\documentclass[%
 reprint,
superscriptaddress,
longbibliography,
showkeys,
 amsmath,amssymb,
 aps,
pre, 
]{revtex4-2}

\usepackage{graphicx}  
\usepackage{dcolumn}   
\usepackage{bm}        
\usepackage{amssymb}   
\usepackage{amsmath}   
\usepackage{mathtools} 
\usepackage{xcolor}
\usepackage{braket}
\usepackage{ulem}
\usepackage[colorlinks,linkcolor=blue,urlcolor=blue,citecolor=blue]{hyperref}
\usepackage{algorithm}
\usepackage{algpseudocodex}

\usepackage{enumitem}
     
\usepackage{dcolumn} 
\usepackage{bm} 

\usepackage{hyperref}
\hypersetup{
    colorlinks=true,
    linkcolor=blue,
    filecolor=magenta,      
    urlcolor=cyan,
    citecolor = purple,
    pdftitle={Overleaf Example},
    pdfpagemode=FullScreen,
    }
\usepackage{array}
\usepackage{float}

\begin{document}

\title{Phase transitions in work fluctuations in correlated N-particle resets}
\author{Felix Herber}
\email{felix.herber@fysik.su.se}
\affiliation{Department of Physics, Stockholm University, 106 91, Stockholm, Sweden \\}
 \author{Deepak Gupta}
\email{phydeepak.gupta@gmail.com}
\affiliation{Institute of Physics, Polish Academy of Sciences, Al. Lotników 32/46, 02-668,
Warsaw, Poland} 
\author{Supriya Krishnamurthy$^{1}$}
\email{supriya@fysik.su.se}

\noaffiliation
\newcommand{\dg}[1]{{\color{red}\bf [DG: #1]}}
\newcommand{\addg}[1]{{\color{red}#1}}
\newcommand{\fh}[1]{{\color{blue}\bf [FH: #1]}}
\newcommand{\sk}[1]{{\color{cyan}\bf [SK: #1]}}
\date{\today}

\begin{abstract}
Systems of $N$ independent uncorrelated particles subjected to the same stochastic change in the environment have recently been studied both theoretically and in experiments, to elucidate the interesting phenomenon of dynamically emergent interparticle positional correlations. However,
the thermodynamic cost of creating such correlations, as well as its interplay with the initial distribution of the $N$ particles has not been studied. In this paper, we study systems of $N$ particles that are simultaneously subjected to a switch of an external potential, either stochastically or periodically, beginning from {\it arbitrarily correlated initial states}. We map out the distribution of thermodynamic work for such a system in the full parameter space of the problem that includes the number of particles $N$, arbitrary initial correlations between particles and arbitrary initial reduced single particle distributions. We demonstrate that the parameter space is partitioned non-trivially, and in an $N$-dependent way, into distinct ``phases", whereby  the functional form of the variance of the work changes discontinuously across phases. The number of such phases depends on the source of stochasticity in the dynamics. 
\end{abstract}
\maketitle

{\it Introduction.---}Stochastic resetting of a single particle or a single degree of freedom has been studied very extensively over the last $15$ years, due to its intrinsic interest as a tractable out-of-equilibrium system, as well as its many applications \cite{evans2020stochastic, Pal_2022, Shamik_2022}.
In recent years, experiments have also been performed that implement single particle resetting \cite{Besga_2020, tal_friedman_2020, Faisant_2021, goerlich2024tamingmaxwellsdemonexperimental}, which corroborate all theoretical results \cite{Besga_2020, tal_friedman_2020, Faisant_2021}.
In these experiments, a colloidal particle moves freely or within a trap until it is reset by turning on an optical confining (resetting) trap that either drags the particle to a predetermined position \cite{tal_friedman_2020} or simply allows the particle to relax to the corresponding equilibrium thermal distribution in the trap \cite{Besga_2020, Faisant_2021, goerlich2024resetting}. 
The confining trap is then turned off and the entire process can  be repeated.
The intervals of time between the turning on of the resetting trap can be stochastically distributed \cite{tal_friedman_2020,goerlich2024tamingmaxwellsdemonexperimental, Faisant_2021} or deterministically periodic \cite{Besga_2020,Faisant_2021}.

Very recently, guided by previous theoretical results \cite{Biroli_2023,Biroli_2024, Biroli_2024L}, an experiment similar to the ones described above has been carried out for investigating multi-particle steady states. Namely $N$ otherwise non-interacting (and initially uncorrelated) particles are simultaneously subjected to a switching trap stiffness~\cite{Biroli_2025}: the particles ($N=4$ in the experiment) diffuse in a trap with a stiffness $k_U$ till the stiffness is instantaneously changed to $k_V$, at which point they diffuse in this new potential before it is switched back to $k_U$. The time intervals between switches are stochastic. 
In this case, it is known from earlier theory~\cite{Biroli_2023,Biroli_2024L,Biroli_2024} that the very interesting phenomenon of dynamically induced interparticle correlations emerges entirely due to the  fact that the $N$ particles are simultaneously subjected to  switching trap stiffnesses at stochastic time intervals \cite{Majumdar_2026}. The results of the experiment matches earlier theoretical predictions for the emergent correlations \cite{Biroli_2025}. Note that the experiment as described above is
more general than the resetting scenario, which is the special case when the particles are allowed to relax to the equilibrium state after the switch to the steeper trap. A different experiment \cite{Yael_2025} carries out the resetting of $N$ particles by dragging them, separately or together, back to their initial positions.

An interesting aspect which these experiments help to quantify, is the idea of assigning a thermodynamic cost to a reset ({\it i.e.} a sudden switching of the trap stiffness), either in terms of the work performed in the implementation of the process or the heat dissipated by the (overdamped) particle once it has been subjected to such a procedure \cite{fuchs2016stochastic,Entropy_UD,Deepak2022_work, mori2023entropy,olsen2024thermodynamic,olsen2024thermodynamic2,gupta2025thermodynamic}. It has been previously shown that these thermodynamic costs can be consistently assigned when $N=1$, for experiments such as those described above, when the resetting  trap  is turned on instantaneously \cite{Deepak2022_work, mori2023entropy,olsen2024thermodynamic} or  over a finite duration \cite{gupta2025thermodynamic}.
\begin{figure}[!t]
    \centering
    \includegraphics[width=0.5\linewidth]{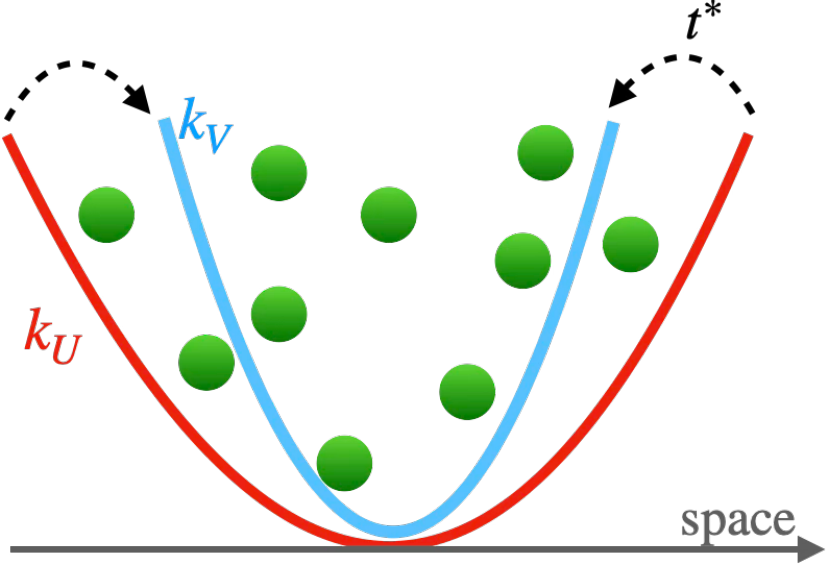}
    \caption{ The schematic represents the class of protocols we investigate. Non-interacting and identical particles are confined in a harmonic trap of stiffness $k_U$. At a time $t^*$, the stiffness is changed instantaneously to $k_V$. We study three cases 1) $t^*$: stochastic and $k_V$: fixed; 2) $t^*$: fixed and $k_V$: random; 3) both $t^*$ and $k_V$ are fixed.}
    \label{fig:scheme}
\end{figure}
A similar analysis has not been performed however for the multi-particle case. In this paper, motivated by earlier work \cite{Biroli_2023, Biroli_2024, Biroli_2024L, Biroli_2025, Majumdar_2026} we study
the scenario when all $N$ particles are subjected to the same dynamics; beginning from an arbitrarily correlated initial state, they diffuse independently in a trap $k_U$
till a time $t^*$ (see Fig. \ref{fig:scheme}) at which point
the trap stiffness is instantaneously changed to $k_V$.
We study the distribution of the work for one such switch of trap stiffness. 

A very important ingredient of our analysis (as opposed to previous work) is including the effect of initial positional  correlations between the $N$ particles. Namely, the particles, could have positional correlations even before the process begins, given by some initial $N$-particle distribution.
All quantities of interest need to be averaged over this fixed initial distribution. Such correlations 
are very relevant when describing  
experiments such as \cite{Biroli_2025} 
where the time intervals  between switching between the two potentials $k_U$ and $k_V$
does not necessarily allow the particles to relax in either, hence resulting in correlated initial states at the end of every switch. Indeed it is shown in \cite{deMauro_2026} that these correlations actually grow in time, reaching eventually a steady-state value. Our analysis would then provide the work cost of one switch with the steady-state distribution as the initial state.
Alternatively one could conceive
of an experiment which brings the particles back to the same correlated initial state after every switch from $k_U$ to $k_V$, by for example, implementing a procedure such as \cite{Sabhapandit_2024} in the $k_V$ trap. 
  
\begin{figure}[t]
    \centering
    \includegraphics[scale=0.12]{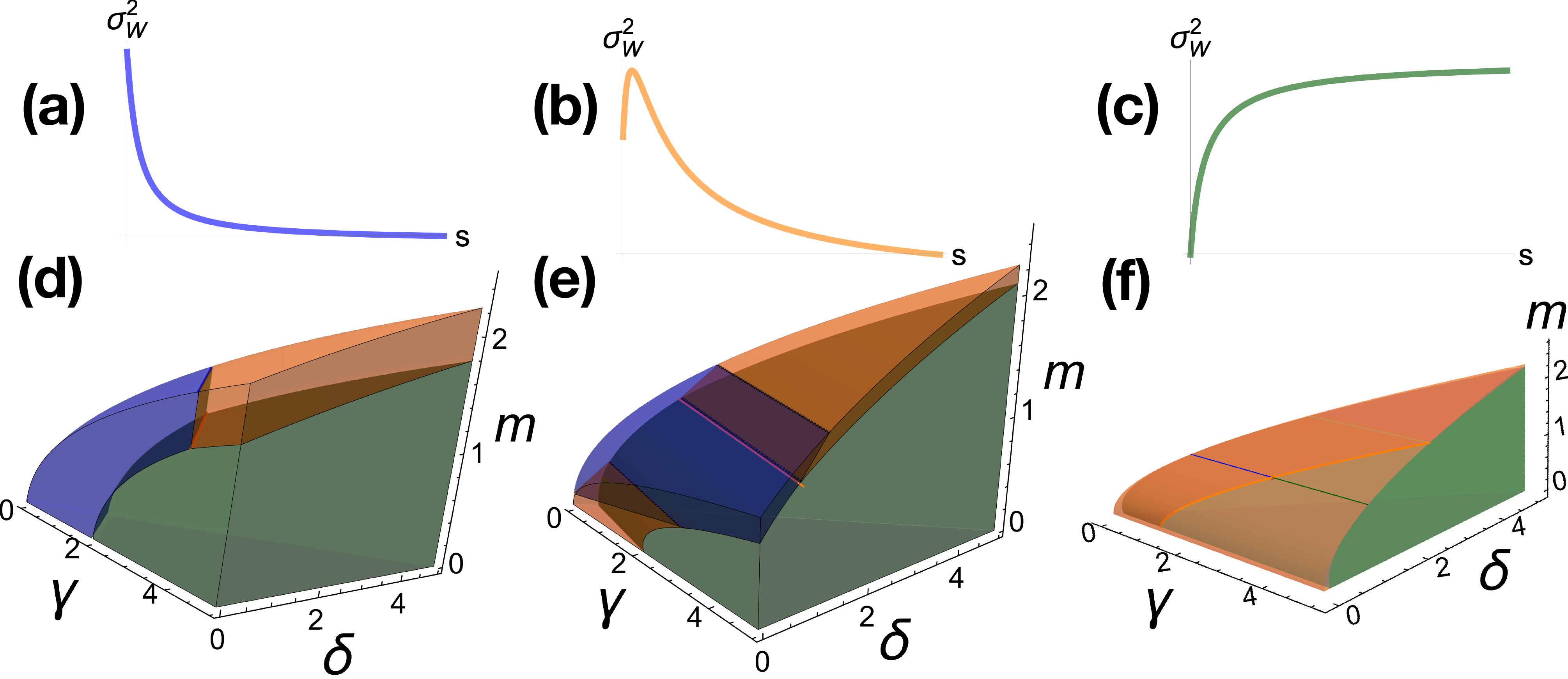}
    \caption{{\it Stochastic $t^*$}. The schematics of different functional forms (a), (b), (c) of the variance of the work $\sigma_W^2$ as a function of the resetting rate. Also shown are the regions occupied by the three ``phases" in the $3$-d parameter space, for d) $N = 2$, e) $N = 7$, and f) $N = \infty$. The blue (a), orange (b) and green (c) curves correspond to 
    the functional form of the variance in the regions in phase space 
    with the same color in panels (d)-(f). The time interval $t^*$ to switching from potential $k_U$ to $k_V$ is picked from the distribution $re^{-rt^*}$ with $s \equiv r/k_U$.}
    \label{fig:Phase_Illustration}
\end{figure}

\begin{figure}
    \centering
    \includegraphics[width=1\linewidth]{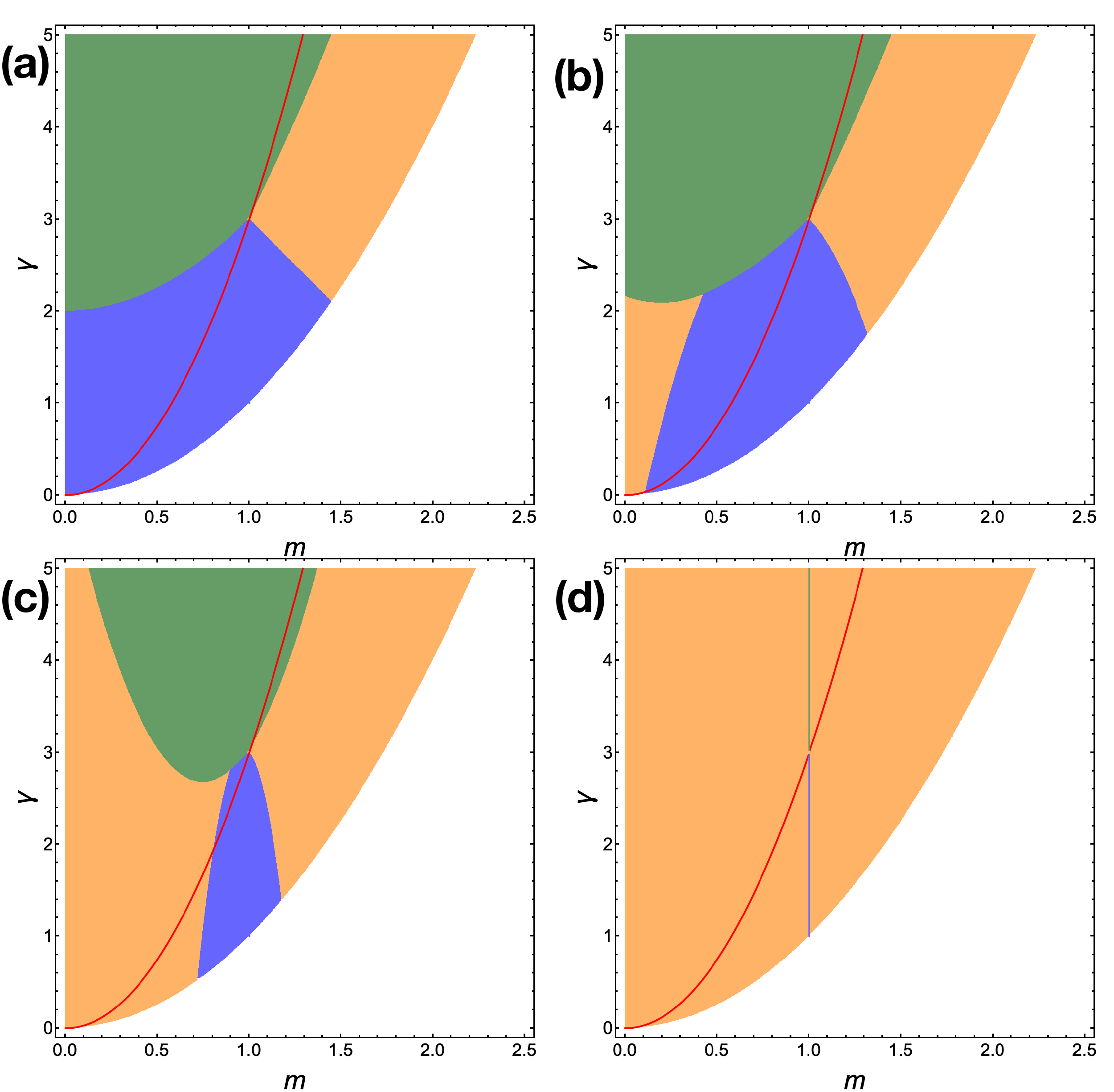}
    \caption{
    {{\it Stochastic $t^*$}. Two-dimensional slice of Fig.~\ref{fig:Phase_Illustration}(d-f) at $\delta = m^2$. The condition $\delta = m^2$ corresponds to the factorized initial $N$-particle states. a) Particle $N = 1$ case (not displayed in Fig.~\ref{fig:Phase_Illustration}) also exhibits a phase structure. Panels b) and c) show $N =7$ and 40~(not displayed in Fig.~\ref{fig:Phase_Illustration}), respectively. d) $N = \infty$: The orange phase occupies the entire region except when $m = 1$ and $\gamma \neq 3$. The red contour $\gamma = 3m^2$ indicates Gaussian initial single particle probability distributions. 
    }}
    \label{fig:StochTimeF}
\end{figure}

\begin{figure}
    \centering
    \includegraphics[width=1\linewidth]{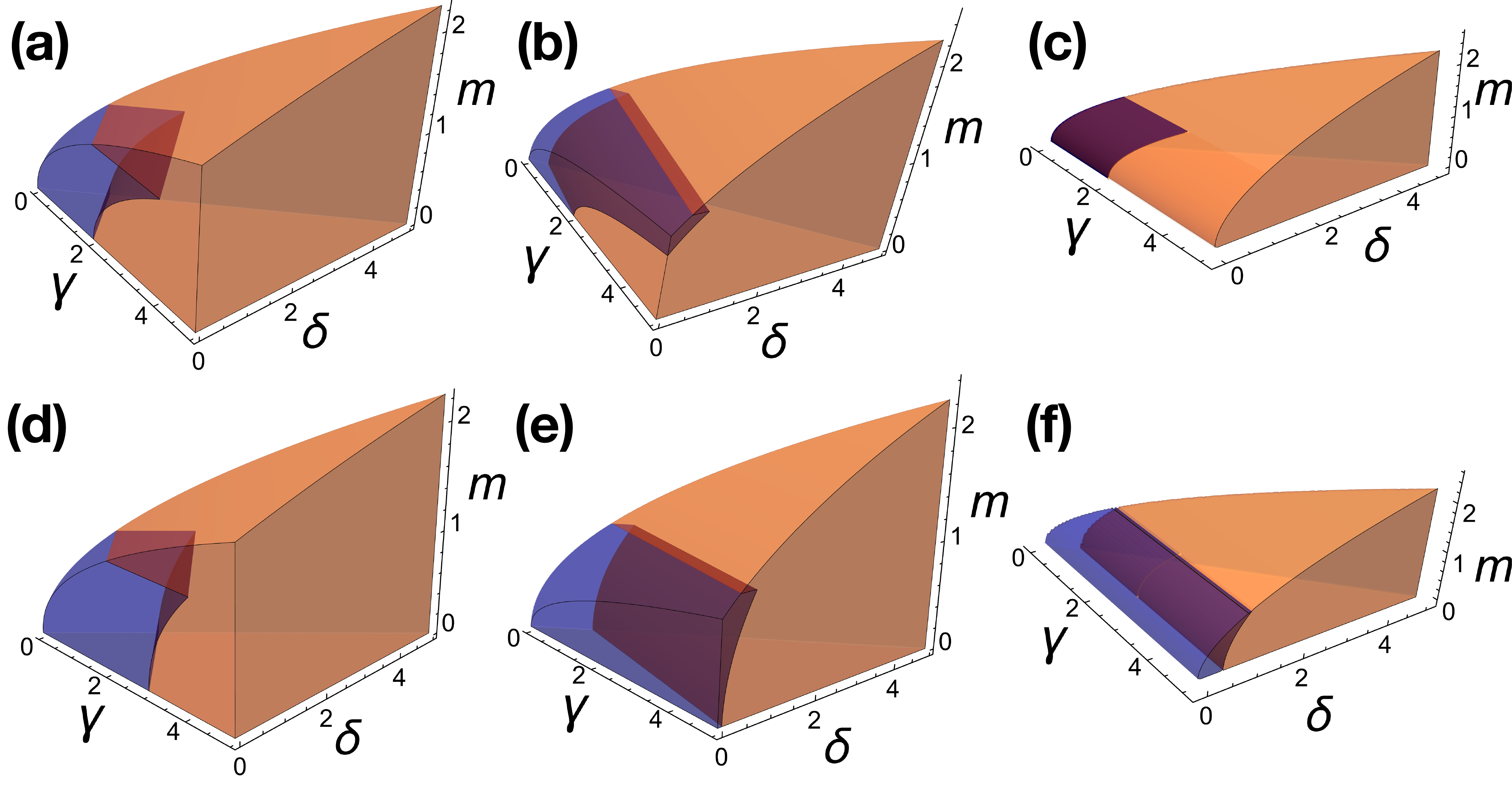}
    \caption{{{\it Deterministic $t^*$}.}
    Phase regions defined by Eq. \eqref{eq:Phase_Random_kV_Final} for $\kappa = 0$ in (a), (b), (c); and $\kappa = \frac{1}{2}$ in (d),~(e), and (f).
    (a) $N = 2$, (b) $N = 7$, (c) $N = \infty$, (d) $N = 2$, (e) $N =7$, and (f) $N = \infty$.}
    \label{fig:PhaseIllustrationsStochKV}
\end{figure}

\begin{figure}
    \centering
    \includegraphics[width = 1\linewidth]{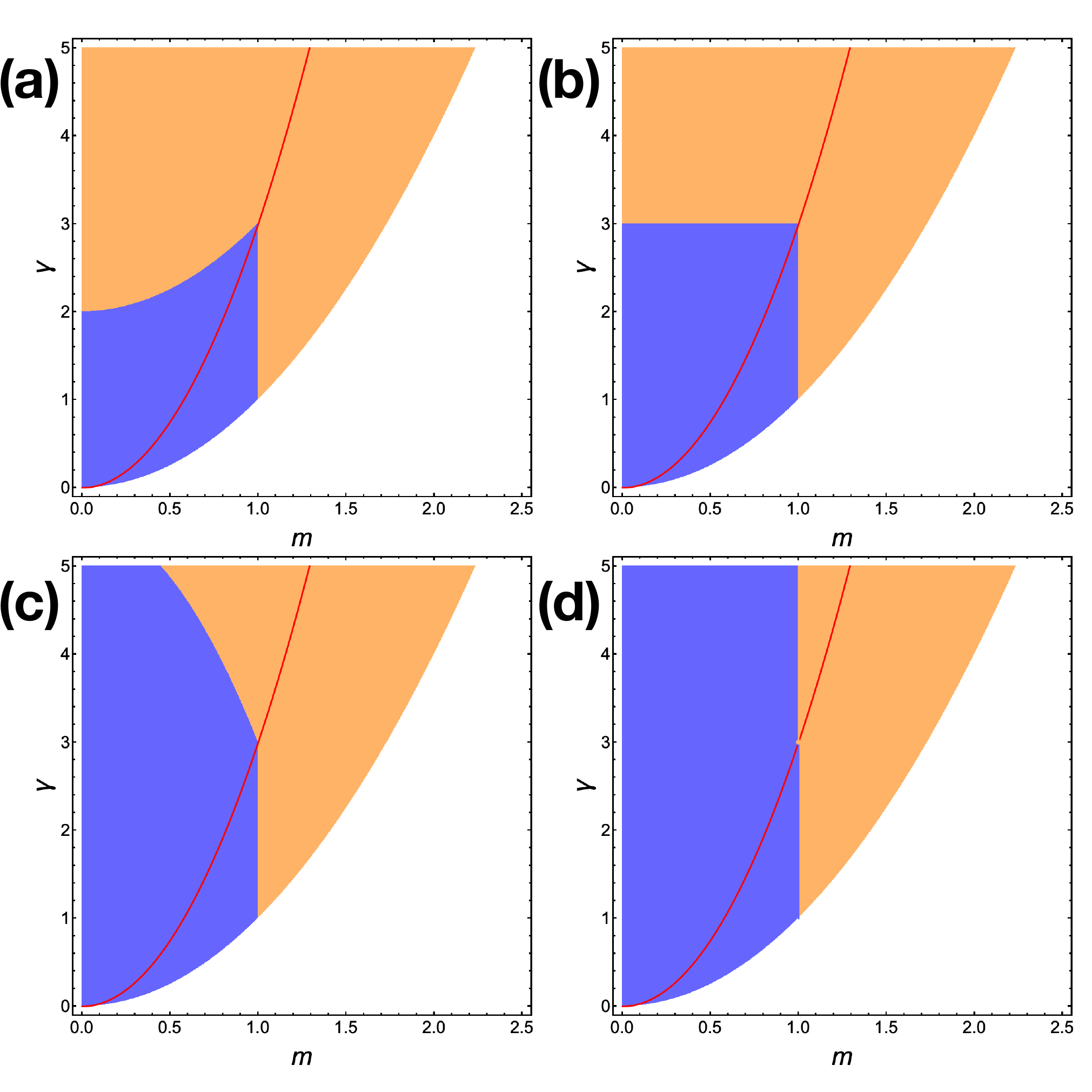}
    \caption{
    {{\it Deterministic $t^*$.} Two-dimensional slice of Fig.~\ref{fig:PhaseIllustrationsStochKV} at $\delta = m^2$, {\it i.e.,} a factorized initial state. Panels (a)-(d) differ by values of $\kappa$ and $N$. (a) $\kappa=0$ case is independent of $N$; (b) $\kappa = \frac{1}{2}$ and $N = 2$; (c) $\kappa = \frac{1}{2}$ and $N = 7$; and (d) $\kappa$ can take any value in $0 <\kappa \leq 1$ and $N \to \infty$. The red contour $\gamma = 3m^2$ corresponds to  a factorized Gaussian initial state. }
    }
    \label{fig:placeholder}
\end{figure}

  We consider three experimentally relevant scenarios: when the trap $k_V$ is switched on at exponentially distributed time intervals (as in \cite{Biroli_2025}), when $k_V$ is switched on at periodic intervals and a third scenario when the value $k_V$ itself is picked from a distribution (for a fixed $t^*$). 
For all these cases, we obtain the full moment generating function of the work distribution, whence we also obtain the mean and variance of the work. We find that though the mean is insensitive to initial interparticle correlations, the variance is very sensitive to the details of the dynamics as well as to details of the initial distribution.
For all $N \geq 1$, the variance of the work shows a phase transition in its functional form (Figs. \ref{fig:Phase_Illustration}, \ref{fig:StochTimeF}, \ref{fig:PhaseIllustrationsStochKV}, and \ref{fig:placeholder})
in the full parameter space of the problem, whereby the position of its global maximum changes. We identify an order parameter related to the position of this maxima. The location of the ``phases'' in the parameter space depends non-trivially on all the parameters, namely the initial interparticle correlations, the initial reduced single particle distribution, the normalized variance of the value of $k_V$ (in the case it is stochastic) and the exact value of $N$ (Figs. \ref{fig:Phase_Illustration}, \ref{fig:StochTimeF}, \ref{fig:PhaseIllustrationsStochKV}, and \ref{fig:placeholder}). The number of phases depends on the source of stochasticity in the problem, changing from $3$ when $t^*$ is stochastic to $2$ when $t^*$ is deterministic 
(with $k_V$ fixed or stochastic).  This rich and surprising phase structure occurs entirely due to the interplay of initial conditions with the out-of-equilibrium dynamics. 

{\it Setup and Results.---}We consider $N$ non-interacting and identical Brownian particles in a single harmonic trap. An equivalent and 
experimentally feasible \cite{Biroli_2025} variant is to consider the $N$ particles to each be in separate but
identical harmonic potentials. The $i$th particle experiences a potential of the form $V_i\big(x_i,t \big) = \frac{k(t)}{2}x_i^2$ with $x_i$ the position of the $i$th particle and $k(t)$ the stiffness of the trap. $k(t)$ is identical for all $N$ particles. The over-damped Langevin equations for these $N$ particles is
    \begin{align}
        \label{eq:Langevin_Equations}
        \frac{\text{d}x_i(t)}{\text{d}t} = -k(t)x_i(t)+\sqrt{2D}\ \eta_i(t), \; i = 1, \dots, N
    \end{align}
    with $\{\eta_i(t)\}_{i = 1}^N$ being independent white Gaussian noises, that is Gaussian stochastic processes with $\braket{\eta_i(t)} = 0$ and $\braket{\eta_i(t_1)\eta_j(t_2)} = \delta_{ij}\delta(t_1-t_2)$, where the angled brackets $\langle \cdots\rangle$ denote the ensemble average over noise realizations.
    In this paper, we will consider the class of protocols that have the form
    \begin{align}
        \label{eq:Protocol_Class}
        k(t) = k_U + \theta(t-t^*)\ (k_V-k_U)\ ,
    \end{align}
    where $k_{U,V} \geq 0$, $ t^* \geq 0$, and the Heaviside step function $\theta(x) =1$ for $x>0$, and $=0$ otherwise.

    The interpretation of this protocol is that the $N$ particles, beginning from some (possibly correlated) initial state,  diffuse simultaneously (and independently of each other) in the potential $k_U$ until the time $t = t^*$ at which point the stiffness of the trap experienced by the particles changes instantaneously to $k_V$. After that, the $N$ particles evolve in the new potential. 
    $k(t)$ could be chosen to be a fully deterministic function where the change from $k_U$ to $k_V$ is always performed at the same time $t^*$ (for this case, an initially factorized state remains factorized \cite{Majumdar_2026}). Alternately,
    stochasticity in $k(t)$ can be introduced in one of two ways, by choosing the time $t^*$ from a distribution (we consider $P(t^*) = re^{-rt^*}$ with $t^* \geq 0$ which is the standard Poissonian case also considered in the experiment \cite{Biroli_2025}) and/or choosing $k_V$ itself from a distribution $P_V(k_V)$ at fixed or random $t^*$. 

    For this system, we are interested in the statistics of the work done by the experimenter on completing one such switch from $k_U$ to $k_V$. The total work performed in a long time series with many such switches can then be pieced together with this information. 
    Since work is only performed at the moment $t^*$ when the potential changes, for the $i$th particle, the work done is simply $W_i (t^*)=  \frac{k_V-k_U}{2}x^2_i(t^*)$, where $x_i(t^*)$ is the position of the $i$th particle, after it has evolved in the $k_U$ potential till the time of the switch $t^*$. The total work done on the $N$ particles is hence $W^{\text{Tot}}_N (t^*)= \sum_{i =1}^N W_i (t^*)$. 
    For arbitrary initial conditions, denoted by
    $\bar{x}_0 = (x_{01}, \dots, x_{0N})$, we can compute the moment generating function of the work 
    $\braket{e^{ -\xi W^{\text{Tot}}_N (t^*)}} $
    (see Appendix A1 for details) to obtain:
    \begin{align}
    \label{eq:Momgen_Function_Setting_N_particles}
 \Phi_N(\omega ;\bar{x}_0|\tau^*) &= \frac{\exp{\Big[-\frac{\omega |\bar{x}_0|^2e^{-2\tau ^*}}{1+2\omega u(\tau^*)}\Big]}}{\sqrt{1+2\omega u(\tau^*)}^N}\ .
    \end{align}
   For convenience, we have defined $u(\tau^*)\equiv 1-e^{-2\tau^*}$ and $\omega \equiv \frac{D}{2k_U}(k_V-k_U) \xi$, where $\xi$ is the Fourier conjugate variable with respect to the work $W^{\text{Tot}}_N$. Furthermore, we have set $\tau^* = k_U t^*$,  
    in addition to transforming to non-dimensionalized coordinates $\bar{x}_0 \to \bar{x}_0\sqrt{D/k_U}$.

     From the moment generating function Eq.~\eqref{eq:Momgen_Function_Setting_N_particles} we can compute any moment of the work. 
     In particular we obtain the first and second moments of the non-dimensionalized work by taking first and second order derivatives with respect to $\omega$.
     We then find the following expressions:
    \begin{align}
        \label{eq:Setting_Work_Before_average}
        \braket{W} &= Nu(\tau^*)+\braket{|\bar{x}_0|^2}e^{-2\tau^*}\ , \\
        \notag
       \braket{W^2} &= 2(N+2)\braket{|\bar{x}_0|^2}e^{-2\tau^*}u(\tau^*)\ + \\
        \label{eq:Setting_2nd_Work_Mom_Before_average}
        &+\braket{|\bar{x}_0|^4} e^{-4\tau^*}+N\big(N+2\big)u(\tau^*)^2\ .
    \end{align}
    The averages indicated by $\langle \cdots \rangle$ on the right-hand side of 
    Eqs.~\eqref{eq:Setting_Work_Before_average} and \eqref{eq:Setting_2nd_Work_Mom_Before_average} are over the initial distribution of the particles. The first and second moments of the work 
    $\braket{W}$ and $\braket{W^2}$ can then be analysed for all the different cases of interest: for $\tau^*$ distributed according to some $P(\tau^*)$, a deterministic $\tau^*$ and for the case where the potential $k_V$ is picked from a distribution $P_V(k_V)$ (see Appendix A2 for more details).
   
   So far we have not assumed anything about the interchangeability of the $N$ particles. For all that follows, we now assume that the $N$ particles are identical. This implies that their initial joint distribution function $P_0 (x_{01}, \dots, x_{0N})$ is exchangeable, {\it i.e.} it is invariant under a permutation of the indices.
In this case, the mean work $\braket{W}$ is simply $N$ times the work for a single particle, since $\braket{|\bar{x}_0|^2} = N 
\braket{x_{01}^2}$. The variance of the work $\sigma_{W^{\text{Tot}}_N}^2 = \braket{W^2}-\braket{W}^2$ becomes
   \begin{align}
       \label{eq:Work_Var_Decomposition}
       \sigma_{W_N^{\text{Tot}}}^2 = N\sigma_{W_1}^2+N(N-1)\text{Cov}(W_1,W_2)\ .
   \end{align}
The term $\text{Cov}(W_1,W_2)$ vanishes when the initial state is factorized and there is no stochasticity in the dynamics. In all other cases it is non-zero. 
Henceforth, we refer to $ \sigma^2_{W_N^{\text{Tot}}}$ as simply $\sigma^2_W$ for ease of presentation.

Equation~\eqref{eq:Work_Var_Decomposition} for the variance carries information about
the dynamics, the initial single particle distributions as well as initial interparticle correlations.  
We analyze  $\sigma^2_W$ for the cases of interest.

{\it Stochastic $\tau^*$.---}We first consider the case where $\tau^*$ is taken from a distribution $P(\tau^*)= r e^{-r\tau^*}$, which is the standard scenario often considered in experiments \cite{Biroli_2025}. To simplify the analysis, we
introduce in Eq. \eqref{eq:Work_Var_Decomposition} the following abbreviations for the initial conditions
$m \equiv \braket{x_{01}^2}$, $\gamma \equiv \braket{x_{01}^4}$, and $\delta \equiv \braket{x_{01}^2x_{02}^2}$. Notice that $\delta > m^2$ ($\delta < m^2$) denotes positively (negatively) correlated initial states, uncorrelated otherwise. We then find for the non-dimensionalized work variance 
   \begin{align}
    \label{eq:Normalized_Work_Var}
    \sigma_{W}^2(s) = \begin{cases}
        0 & s= 0\\
        2N+\frac{s}{s+4}a_N+\frac{s}{s+2}b_N \\
        -\frac{s^2}{(s+2)^2}(m-1)^2N^2 &s >0 \\
    \end{cases}\ .
    \end{align}
    Here, $s\equiv r/k_U$ and we have introduced $a_N \equiv (\gamma-\delta -4m+2)N+(\delta-2m+1)N^2$ and $b_N \equiv 4N(m-1)$.
    The mean work as  a function of $s$ depends on the initial conditions only through $m$ and shows no change in behavior in different regions of the phase space. However,
    $\sigma_{W}^2(s)$ exists in three distinct  ``phases'' depending on the initial conditions $(m, \gamma, \delta)$. 
    These phases are distinguished by the location at which $\sigma_{W}^2(s)$ attains its supremum: in the limit $s\to 0$ (Phase B, blue in all the figures), at a finite value of $s>0$ (Phase O, orange), or in the limit $s\to\infty$ (Phase G, green). Hereafter, we refer to these phases as the blue-, orange-, and green-phases, respectively.
    It is possible to show (see Appendix B1 for more details) that for Eq. \eqref{eq:Normalized_Work_Var} the orange-phase will occur for some initial conditions $(m, \gamma, \delta)$ and finite $N$, if and only if
    \begin{align}
        \label{eq:Tot_Work_Var_Phase_1}
        \begin{cases}
            \gamma +(N-1)\delta = N+2 &m=1, \\
            -2B_N<A_N\leq \frac{(B_N-3)^2}{8} &m\neq 1, \;  -1 < B_N < 1 \\
            -2B_N<A_N<1-\frac{B_N}{2} &m\neq 1, \quad B_N \geq 1
        \end{cases}\ .
    \end{align}
   Here $A_N \equiv \frac{a_N}{N^2(m-1)^2}$ and $B_N \equiv\frac{4}{N(m-1)}$, where $a_N$ is given below Eq.~\eqref{eq:Normalized_Work_Var}. 
   Similar conditions exist for each of the other phases (see Appendix B1 for details). The analysis
   leading to these conditions is involved because of the many conditions imposed on the moments by the initial many-particle and reduced single-particle probability distributions (Appendix C1). In addition special cases, such as the equilibrium case $m=1$ as well as when $N = \infty$ need to be accounted for separately. Appendix B1 details these calculations. The three distinct functional forms of the variance are shown in Fig.~\ref{fig:Phase_Illustration}, which also shows the 
regions occupied by the three phases in the parameter space $(m, \gamma, \delta)$ for different $N$. 

 Each allowed value of $(m, \gamma, \delta)$ characterizes an initial probability distribution of $N$-particles.
A particularly simple but relevant initial distribution is a factorized one, corresponding to $\delta = m^2$. The single particle distributions in the factorized product are specified by the specific values of $\gamma$ and $m$. We find from Eq.~\eqref{eq:Normalized_Work_Var} that the work variance with non-factorized and positively correlated initial states is always larger than for factorized states. 

Figure~\ref{fig:StochTimeF} shows a $2$-d section of the $3$-d regions in Fig.~\ref{fig:Phase_Illustration}. 
The red contour $\gamma = 3m^2$ depict Gaussian single particle distributions. 
Such an initially factorized distribution of Gaussians would naturally arise if 
the initial distribution was generated by equilibrating all $N$ particles in the $k_V$ trap. This case has been studied in both theory and experiments, for $N=1$, in \cite{Besga_2020,Faisant_2021,olsen2024thermodynamic}. As can be seen, when $N=1$ and $\gamma=m^2$, the variance lies exclusively in the blue-phase for $m<1$ and in the green-phase for $m>1$. This seems to be the case even for low values of $N \lesssim 5$ (see Appendix C2), making it easy to miss this phase space structure in earlier studies. The point $m=1,\gamma = 3$ is a special point corresponding to initial states that are equilibrium
states in the $k_U$ trap. This state is always in the orange phase irrespective of $N$ because the variance of the work is a constant, independent of $s$. 

In Appendix D1, we analyze Eq.~(\ref{eq:Normalized_Work_Var}) further to understand how the variance changes non-analytically across the different phase boundaries. Very interestingly, we find that the kind of non-analyticity depends on which of the $3$ boundaries is being traversed. Across the orange-green as well as the blue-orange phase boundaries, the order parameter behaves as for a second-order phase transition, whereas across the blue-green phase boundary, it is like a first-order phase transition.

{\it Fixed $\tau^*$~+~stochastic $k_V$.---}We now consider $\tau^*$ to be fixed but $k_V$ to be stochastic. We first multiply Eq.~\eqref{eq:Setting_Work_Before_average} by the factor $\frac{1}{2}(k_V-k_U)$ and Eq.~\eqref{eq:Setting_2nd_Work_Mom_Before_average} by the factor $\frac{1}{4}(k_V-k_U)^2$ before taking the average with respect to $k_V$.
We find that 
the non-dimensionalized work variance is

\begin{widetext}
    \begin{equation}
\begin{aligned}
\sigma_W^2 =&
\frac{N}{4}\langle (k_V-k_U)^2\rangle
\Big[
(N+2)\mathcal{A}^2+
\Big(\gamma+(N-1)\delta-(N+2)m^2\Big)
e^{-4\tau^*}
\Big]-\frac{N^2}{4}
\big(\langle k_V\rangle-k_U\big)^2 \mathcal{A}^2\ ,
\end{aligned}
\end{equation}
    \label{eq:Work_Var_Random_kV}
\end{widetext}
where $\mathcal{A}\equiv u(\tau^*)+m~e^{-2\tau^*}.$
Along with the parameters ($m,\delta,\gamma, N$) we had before, we now also have 
$\braket{k_V}, \braket{k_V^2}$ which are the first two moments of the chosen distribution $P_V(k_V)$. If we consider these two parameters as fixed, then one can show that $\sigma^2_{W}$ will attain its supremum for finite $\tau^*$ if and only if one of the conditions below hold
\begin{align}
     \label{eq:Phase_Random_kV_Final}
     \begin{cases}
        m > 1, \\
       C_N \geq 2+N(1-m^2)\ \kappa
    \end{cases}.
\end{align}
Here $C_N \equiv \gamma-m^2+(N-1)(\delta-m^2)$; and $\kappa \equiv \frac{\braket{\Delta k^2}-\braket{\Delta k}^2}{\braket{\Delta k^2}}$ where $\Delta k \equiv k_V-k_U$. There are only two phase space regions in this case, illustrated in Fig.~\ref{fig:PhaseIllustrationsStochKV}.  These are specified  with respect to whether the variance of the work attains its supremum at a finite $\tau^*$ (orange) or infinite $\tau^*$ (blue)~\footnote{The color coding is same as in the stochastic $t^*$ case for finite $s>0$  (orange) and $s\to 0$  (blue).}. 
Appendix B2 details the conditions for the two phases for finite and infinite $N$. Figures~\ref{fig:PhaseIllustrationsStochKV} and \ref{fig:placeholder} illustrate that, apart from the number of phases, the phase space partitioning is also very different  from Figs.~\ref{fig:Phase_Illustration} and \ref{fig:StochTimeF}.

Figure~\ref{fig:placeholder} shows the analog of  
Fig.~\ref{fig:StochTimeF} for the factorized initial states. 
Note that $\kappa = 0$ corresponds to the case when $k_V$ is fixed as also $\tau^*$. As Fig.~\ref{fig:placeholder}a shows, the phase space partitioning in this case is independent of $N$, in keeping with earlier insights that no interparticle correlations are dynamically generated in this case \cite{Majumdar_2026}. The case $\kappa = 1/2$ corresponds to the case $k_V$ is stochastic (with specified mean and variance) and deterministic $\tau^*$. Stochasticity in $\tau^*$ can easily be added here too, but we make this choice for displaying the contrast between stochasticity only in $\tau^*$ and stochasticity only in $k_V$.  In Appendix D2 we analyze  the stochastic $k_V$ case further to  understand how the variance changes non-analytically across the phase boundary. We find that the relevant order parameter~(crossing the blue-orange boundary) changes here as in a second-order phase transition similar to stochastic $t^*$ case.

{\it Discussion.---}We have looked at systems of $N$-particles that evolve in a potential $k_U$ until the potential abruptly changes to $k_V$. Such setups for a single particle have been experimentally and theoretically investigated in the different contexts of stochastic resetting \cite{Besga_2020,Faisant_2021,goerlich2024tamingmaxwellsdemonexperimental, gupta2020stochastic, gupta2021resetting}, dynamics in fluctuating potentials \cite{alston2022non,mercado2020intermittent, mercado2022reducing, santra2021brownian} and thermodynamic costs for resetting \cite{mori2023entropy,olsen2024thermodynamic,olsen2024thermodynamic2}. More recently the multi-particle case has also been of interest \cite{Biroli_2023,Biroli_2024L, Biroli_2025,Yael_2025} for the  dynamical correlations that emerge purely through the simultaneous influence on all the particles of the switching trap stiffness \cite{Majumdar_2026}. It is this multi-particle case that we analyze, including in our analysis the effect of arbitrary initial $N$-particle correlations (relevant in describing experiments such as \cite{Biroli_2025}), parametrized by the values of $m,\gamma, \delta$. Our study shows the emergence of a very rich phase space structure which differentiates between the sources of stochasticity, either due to $\tau^*$ or $k_V$. It is possible that looking at the full work distribution may partition the phase space even  further. Moreover, it would also be interesting to calculate the large deviation function corresponding to the work fluctuations for a series of such switches in finite-long time limit, taking into account the (possibly) different correlations after a switch from $k_U$ to $k_V$ or the reverse.

The multi-particle case becomes more complicated if the particles are correlated at all times. This  indeed is the case in the experimental studies \cite{Biroli_2025, Yael_2025}, due to hydrodynamic correlations being present at all times, though intriguingly it was found \cite{Biroli_2025} that the hydrodynamic correlations seemed to have a negligible effect on the interparticle correlations $\delta-m^2$. Since $\delta-m^2$ is also one of the relevant quantities in our calculations of the variance of the thermodynamic work, one might expect the hydrodynamic correlations to not affect our results. However all our results are contingent on being able to write the total work for $N$ particles as a sum of the individual work for each particle. It would be very interesting if a similar analysis is possible even when this is not the case.
Finally, in recent work \cite{olsen2026}, a general bound has been derived for the variance of the work for the stochastic $t^*$ case,  beginning from factorized initial conditions. It would be very interesting to understand if the phase structure that we find for this problem affects such bounds.

{\it Acknowledgments.---}The authors thank Satya N. Majumdar and  Gabriele de Mauro for  carefully reading the manuscript and for providing valuable comments and suggestions that helped improve the draft. The authors would also like to thank
Sergio Ciliberto for fruitful discussions.
D.G. acknowledges the financial support provided by NORDITA and Stockholm University for the academic visit in April 20-28, 2026, during which the collaboration resulting in this project started. D.G. acknowledges support from the Alexander von Humboldt foundation.
S.K. acknowledges the support of the Swedish Research Council through the grant 2021-05070.

\newpage

\appendix
\onecolumngrid
\section{Work variance and moment generating function}
\subsection{Moment generating function derivation}

We want to obtain the moment generating function (MGF) of the work Eq. (\ref{eq:Momgen_Function_Setting_N_particles}) of the main article. We obtain the MGF by first recalling that when the protocol $k(t)$ is deterministic, the diffusion of the $i$th particle is described by overdamped Langevin equation
\begin{align}
    \label{eq:Langevin_Eq}
    \frac{\text{d}x_i}{\text{d}t}(t) = -k(t)x_i(t)+\sqrt{2D}\eta_i(t)
\end{align}
with $\braket{\eta_i(t)} = 0$, $\braket{\eta_i(t)\eta_j(t')} = \delta_{ij}\delta(t-t')$ and $k(t) = k_U+(k_V-k_U)\theta(t-t^*)$. We assume that $t_f$ is the time the experiment runs over, and in that time, there is either no switch at all or one switch to $k_V$. In the main text, we have taken $t_f \rightarrow \infty$ so that there is always one switch.

The work done on the $i$th particle by the experimenter who controls $k(t)$ is $W_i(t_f) =  \frac{k_V-k_U}{2}x^2_i(t^*)\theta(t_f-t^*)$. The MGF of the work performed on particle $i$ is then
\begin{align}
   \label{eq:MGF_Work_Single_Particle_Work}
    \braket{e^{-\xi W_i(t_f)}} = \int_{-\infty}^\infty \text{d}x_i^* \; e^{-\frac{\xi(k_V-k_U)}{2}(x_i^*)^2\theta(t^*-t_f)}P(t^*,x_i^*|x_{0i})
\end{align}
where $P(t^*, x^*|x_{0i})$ is the probability for the particle to be at position $x^*$ at time $t^*$ given that it was at $x_{0i}$ at time $t = 0$.

Eq. \eqref{eq:MGF_Work_Single_Particle_Work} tells us that only the position distribution of the particle at time $t^*$ matters. So we only need to solve eq. \eqref{eq:Langevin_Eq} when $t \leq t^*$. We can therefore take $k(t) = k_U$ in the Langevin equation Eq. \ref{eq:Langevin_Eq}.
\begin{align}
    \label{eq:Langevin_Eq_Below_t_start}
    \frac{\text{d}x_i}{\text{d}t}(t) = -k_Ux_i(t)+\sqrt{2D}\eta_i(t), \;  t\leq t^*.
\end{align}
The Fokker-Planck equation corresponding to eq. \eqref{eq:Langevin_Eq_Below_t_start} is 
\begin{align}
    \label{eq:Fokker_Planck_Eq}
    \frac{\partial P}{\partial t}(x_i^*, t|x_{0i}) = k_U\frac{\partial}{\partial x_i^*}\Big(x_i^*P(x_i^*,t|x_{0i})\Big)+D\frac{\partial^2 P}{\partial x_i^{*2}}(x_i^*,t|x_{0i}), \; \text{ for } t \leq t^*.
\end{align}
The solution to eq. \eqref{eq:Fokker_Planck_Eq} is
\begin{align}
    \label{eq:Solution_Fokker_Planck_Eq}
    P(x_i^*,t|x_{0i}) = \frac{1}{\sqrt{2\pi D(1-e^{-2k_Ut})}}\exp\big(-\frac{(x_i^*-x_{0i}e^{-2k_Ut})^2}{2D(1-e^{-2k_Ut})} \big).
\end{align}
If we substitute eq. \eqref{eq:Solution_Fokker_Planck_Eq} into eq. \eqref{eq:MGF_Work_Single_Particle_Work} and integrate, we find
\begin{align}
    \label{eq:Momgen_Function}
    \Phi(\xi,t_f;x_{0i}|t^*) &= \begin{cases}
        1, \; &0 \leq t_f < t^* \\
        \frac{\exp{\Big[-\frac{\xi (k_V-k_U)x_{0i}^2e^{-2k_Ut^*}}{2[1+2\xi(k_V-k_U)v(t^*)]}\Big]}}{\sqrt{1+2\xi(k_V-k_U)v(t^*)}}, \; &t_f\geq t^*
    \end{cases}
    \\
 \notag
    v(t^*) &= \frac{D}{2k_U}(1-e^{-2k_Ut^*})
\end{align}
All this has just been for one particle. Consider now $N$ independent particles, with the $i$th particle at $x_{0i}$ at $t=0$. We assume also that $k(t)$ is deterministic initially, such that $t^*$ is fixed. In this case, the MGF for the whole system is just a product over the MGF of the $N$ different particles $\Phi(\xi,t;x_{0i}|t^*)$ since then $\Phi(\xi, t_f; \bar{x}_0|t^*) = \prod_{i = 1}^N\braket{e^{-\xi W_i}} = \braket{e^{-\xi \sum_{i = 1}^N W_i}}$ where we in the last equality have used independence of the $N$ particles. Applying these steps to eq. \eqref{eq:Momgen_Function} we find
\begin{align}
    \label{eq:Momgen_Function_N_particles}
    \Phi_N(\xi,t_f;\bar{x}_0|t^*) &= \begin{cases}
        1, \; &0 \leq t_f < t^* \\
        \frac{\exp{\Big[-\frac{\xi (k_V-k_U)|\bar{x}_0|^2e^{-2k_Ut^*}}{2[1+2\xi(k_V-k_U)v(t^*)]}\Big]}}{\sqrt{1+2\xi(k_V-k_U)v(t^*)}^N}, \; &t_f\geq t^*
    \end{cases}
    \\
 \notag
    v(t^*) &= \frac{D}{2k_U}(1-e^{-2k_Ut^*})
\end{align}
Here we have introduced $\bar{x}_0 = (x_{01}, \dots, x_{0N})$ as a short hand for the initial positions of the $N$ particles. 

Eq. (\ref{eq:Momgen_Function_Setting_N_particles}) in the paper can be obtained from eq. \eqref{eq:Momgen_Function_N_particles} by non-dimensionalizing it. To obtain the moments from this expression, note that further averages need to performed with respect to $\bar{x}_0$ and $t^*$ (or $k_V$) depending on the kind of stochasticity we introduce.\\

\subsection{Work variances}
To obtain the first and second moments of the work from Eq. \eqref{eq:Momgen_Function_N_particles}, we differentiate with respect to $\xi$ the appropriate number of times and average over the initial distribution $P_0(\bar{x}_0)$ to get
\begin{align}
    \label{eq:Work_Average_Pos}
    \braket{W} &= -(k_V-k_U)\Big(Nv(t^*)+\frac{1}{2}\braket{|\bar{x}_0|^2}e^{-2k_Ut^*} \Big)\theta(t_f-t^*), \\
    \label{eq:2nd_Work_Mom_Average_Pos}
    \braket{W^2} &= 2(k_V-k_U)^2\Bigg(\braket{|\bar{x}_0|^2}e^{-2k_Ut^*}v(t^*)+\frac{1}{8}\braket{|\bar{x}_0|^4} e^{-4k_Ut^*}+N(\frac{N}{2}+1)v(t^*)^2+\frac{N}{2}\braket{|\bar{x}_0|^2}e^{-2k_Ut^*}v(t^*) \Bigg)\theta(t_f-t^*).
\end{align}
Equations \eqref{eq:Work_Average_Pos} and \eqref{eq:2nd_Work_Mom_Average_Pos} still assume that $k_V$ and $\tau^*$ are fixed. If we want to treat either (or both) of them as stochastic, we just multiply equations \eqref{eq:Work_Average_Pos} and \eqref{eq:2nd_Work_Mom_Average_Pos} by their desired probability distributions $P_{V}(k_V)$ and $P_{\tau^*}(\tau^*)$ and then integrate to get the final values of the moments.

In this paper, when we consider the time $t^*$ to be stochastic, we consider the distribution $P_{\tau^*}(\tau^*) = re^{-r\tau^*}$. Multiplying equations \eqref{eq:Work_Average_Pos} and \eqref{eq:2nd_Work_Mom_Average_Pos} by $P_{\tau^*}(\tau^*) = re^{-r\tau^*}$, doing the integral $\int_0^\infty \text{d}t^*$, and then using the results to compute the work variance $\sigma_W^2 = \braket{W^2}-\braket{W}^2$ we find,
\begin{align}
    \notag
    \sigma_W^2 &= \frac{(k_V-k_U)^2}{4}\Bigg\{\frac{D^2}{k_U^2}N(N+2)(1-e^{-rt_f})\\
    \notag
    &+\frac{2r}{r+2k_U}\Big(\frac{(N+2)D}{k_U}\braket{|\bar{x}_0|^2}-\frac{D^2}{k_U^2}N(N+2)\Big)\big[1-e^{-(r+2k_U)t_f}\big]+\\
    \notag
    &+\frac{r}{r+4k_U}\Big(\braket{|\bar{x}_0|^4}-\frac{2(N+2)D}{k_U}\braket{|\bar{x}_0|^2}+N(N+2)\frac{D^2}{k_U^2} \Big)\big[1-e^{-(r+4k_U)t_f}\big]+\\
    \label{eq:Work_Variance_Stoch_Time}
    &-\Big[\frac{ND}{k_U}[1-e^{-rt_f}]+(\braket{|\bar{x}_0|^2}-\frac{ND}{k_U})\frac{r}{r+2k_U}[1-e^{-(r+2k_U)t_f}] \Big]^2\Bigg\}.
\end{align}
When we consider $k_V$ to be stochastic, we consider an average over an arbitrary distribution $P_{k_V}(k_V)$. Multiplying equations \eqref{eq:Work_Average_Pos} and \eqref{eq:2nd_Work_Mom_Average_Pos} by $P_{k_V}(k_V)$, doing the integral $\int_{-\infty}^\infty \text{d}k_V$, and then using the result to compute the work variance $\sigma_W^2 = \braket{W^2}-\braket{W}^2$ we find,
\begin{align}
    \notag
    \sigma_W^2 = 2\braket{(k_V-k_U)^2}&\Bigg(\braket{|\bar{x}_0|^2}e^{-2k_Ut^*}v(t^*)+\frac{1}{8}\braket{|\bar{x}_0|^4} e^{-4k_Ut^*}+N(\frac{N}{2}+1)v(t^*)^2+\frac{N}{2}\braket{|\bar{x}_0|^2}e^{-2k_Ut^*}v(t^*) \Bigg) + \\
    \label{eq:Work_Variance_Stoch_kV}
    -(\braket{k_V}-k_U)^2&\Bigg(Nv(t^*)+\frac{1}{2}\braket{|\bar{x}_0|^2}e^{-2k_Ut^*} \Bigg)^2.
\end{align}
Here, we have introduced $\braket{(k_V-k_U)^2} = \int_{-\infty}^\infty \text{d}k_V \; P_{k_V}(k_V) (k_V-k_U)^2$ and $\braket{k_V} = \int_{-\infty}^\infty \text{d}k_V \; P_{k_V}(k_V) k_V$. 
\section{Phase Analysis}
Here we will analyze equations \eqref{eq:Work_Variance_Stoch_Time} and \eqref{eq:Work_Variance_Stoch_kV} and check for the conditions for the different phases. We first write Eqns. \eqref{eq:Work_Variance_Stoch_Time} and \eqref{eq:Work_Variance_Stoch_kV} in non-dimensional variables. We do that by first defining variables {\bf s, m , $\gamma$ and $\delta$ } by the relations $r = sk_U$, $\braket{|\bar{x}_0|^2} = \frac{DNm}{k_U}$, $\braket{|\bar{x}_0|^4} = \frac{D^2}{k_U^2}(N\gamma+N(N-1)\delta )$ and $\braket{x_{0i}^2x_{0j}^2} = \frac{\delta D^2}{k_U^2}$ (whenever $i \neq j$), $\tau_f = k_Ut_f$ and lastly $\tau^* = k_Ut^*$. In the case of eq. \eqref{eq:Work_Variance_Stoch_Time} we divide then both sides by $\frac{D^2(k_V-k_U)^2}{4k_U^2}$ and take $\tau_f \to \infty$. In the case of eq. \eqref{eq:Work_Variance_Stoch_kV} we divide both sides by $\frac{D^2}{k_U^2}$. Note we have already used the fact that the initial distribution is exchangeable (equivalent to assuming that the $N$ particles are independent and identically distributed).  
We check if Eq.  \eqref{eq:Work_Variance_Stoch_Time} (and Eq. \eqref{eq:Work_Variance_Stoch_kV}) attain their  suprema for finite $s$ (finite $t^*$ respectively) for the allowed values of the variables m , $\gamma$ and $\delta$ . 
\subsection{Stochastic time}
Using eq. \eqref{eq:Work_Variance_Stoch_Time}, we find for the non-dimensionalized work variance
\begin{align}
    \label{eq:Stoch_Time_Work_Var_Nondim}
    \mathcal{V}(s) 
    &= N\Big(2+\frac{4s}{s+2}(m-1)+\frac{s}{s+4}(\gamma-\delta -4m+2)\Big) +\frac{s}{s+4}N^2\Big(\delta-2m+1\Big)-N^2(m-1)^2\frac{s^2}{(s+2)^2}.
\end{align}
This is just eq. (7) of the main article but written out in full form. 

\begin{figure}[htbp]
    \centering
    \begin{minipage}[b]{0.35\textwidth}
        \centering
        \includegraphics[width=\linewidth]{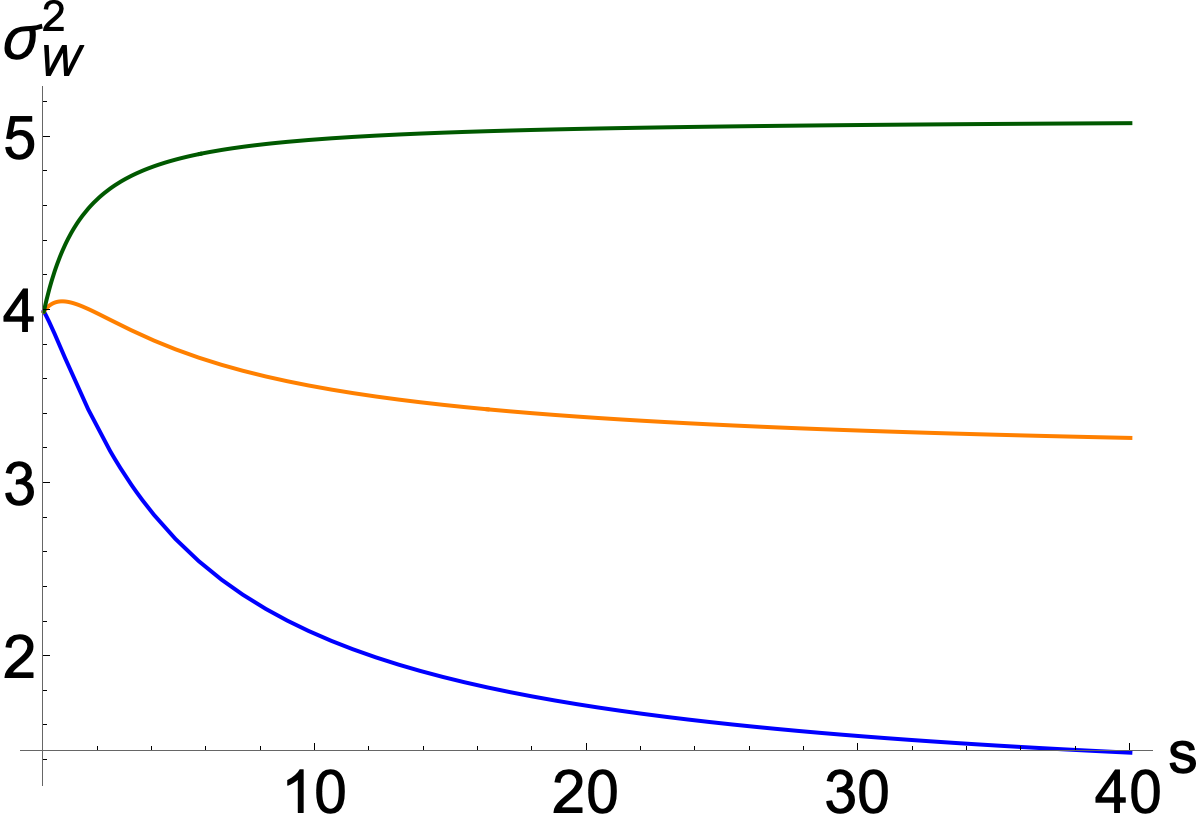}
        \caption{Here, we show plots of how eq. \eqref{eq:Stoch_Time_Work_Var_Nondim} could look like as a function of $s$ if one chooses either $(m, \gamma, \delta) = (\frac{6}{5}, 2, \frac{36}{25})$ in Phase B, $(m, \gamma, \delta) = (\frac{6}{5}, 3, \frac{36}{25})$ in Phase O or $(m, \gamma, \delta) = (\frac{6}{5}, 4, \frac{36}{25})$ in Phase G. These are drawn in blue, orange and green respectively. $N = 2$. }
        \label{fig:StochTimePhaseFunction1}
    \end{minipage}
    \hspace{1cm}
    \begin{minipage}[b]{0.35\textwidth}
        \centering
        \includegraphics[width=\linewidth]{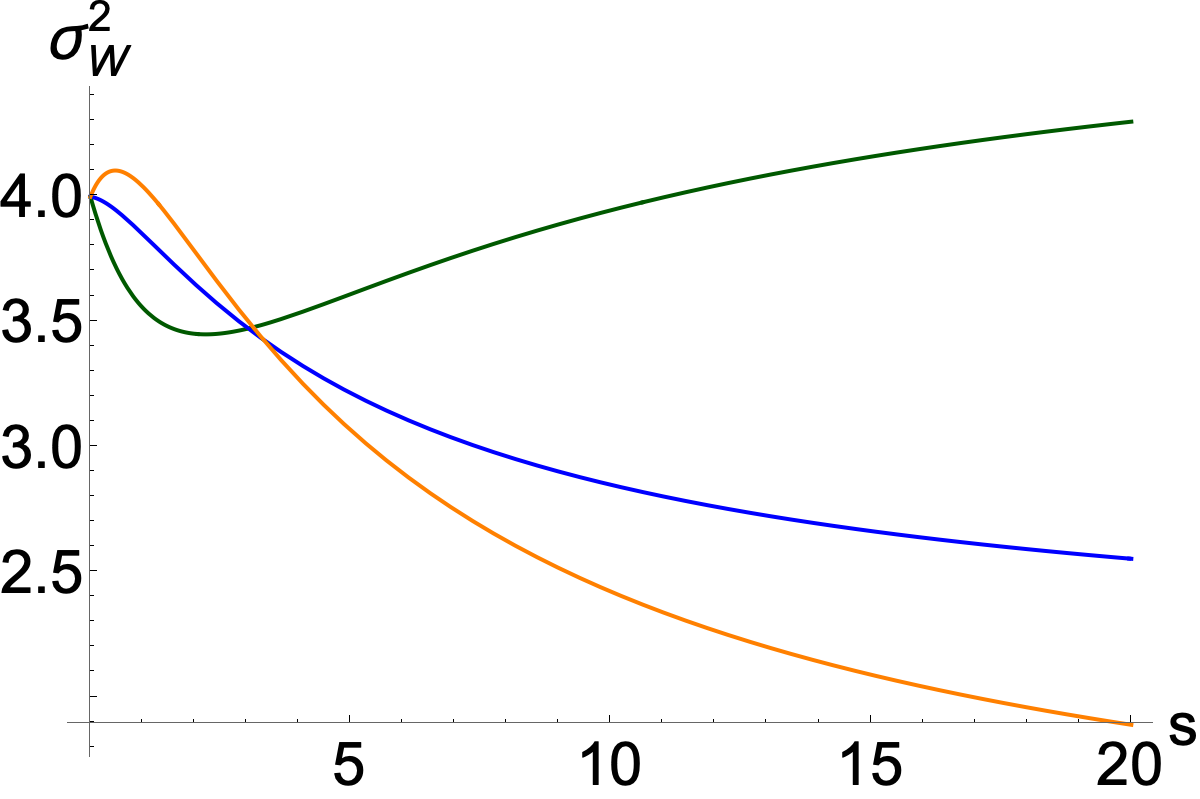}
        \caption{An illustration of how eq. \eqref{eq:Stoch_Time_Work_Var_Nondim} could look like as a function of $s$ if one chooses either $(m, \gamma, \delta) = (\frac{6}{5}, \frac{5}{2}, \frac{36}{25})$ in Phase B. $(m, \gamma, \delta) = (\frac{3}{2}, \frac{5}{2}, \frac{9}{4})$ in Phase O or $(m, \gamma, \delta) = (\frac{1}{5}, \frac{5}{2}, \frac{1}{25})$ in Phase G. These are drawn in blue, orange and green respectively. $N = 2$.}
        \label{fig:StochTimePhaseFunction2}
    \end{minipage}
\end{figure}

\begin{figure}
    \centering
    \includegraphics[width=0.7\linewidth]{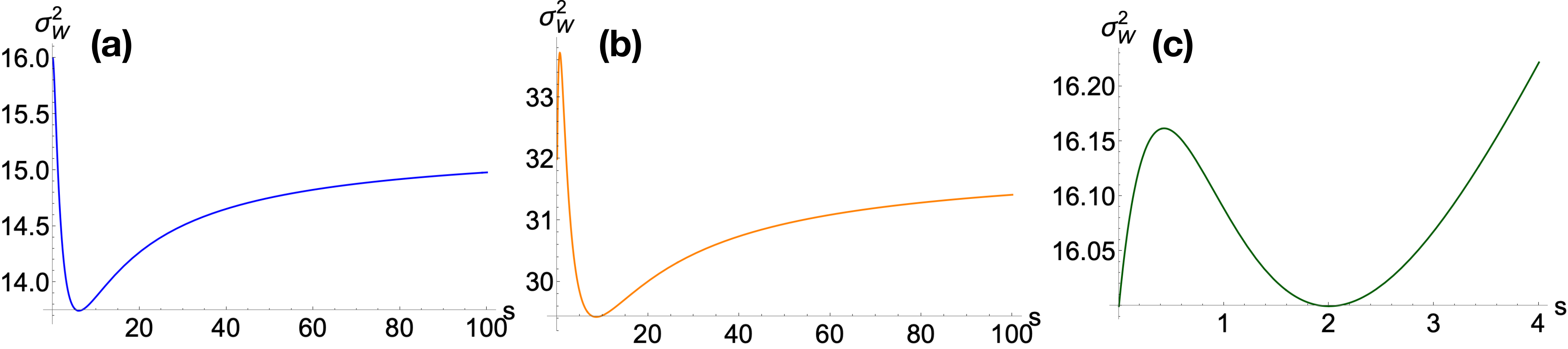}
    \caption{This figures shows how eq. \eqref{eq:Stoch_Time_Work_Var_Nondim} could look like as a function of $s$ if one chooses either (a) $(m, \gamma, \delta) = (\frac{1}{2}, \frac{125}{256}, \frac{125}{256})$ with $N = 8$ in Phase B (b) $(m, \gamma, \delta) = (\frac{1}{2}, \frac{3}{8}, \frac{3}{8})$ with $N = 16$ in Phase O or (c) $(m, \gamma, \delta) = (\frac{1}{2}, \frac{9}{16}, \frac{9}{16})$ with $N = 8$ in Phase G. It turns out that one cannot make eq. \eqref{eq:Stoch_Time_Work_Var_Nondim} have more than two stationary points in the region $s > 0$. It is also not possible to make $\mathcal{V}(s)$ to have more than one stationary point if one wants the supremum to be approached as $s \to 0^+$ and not attained for any finite $s >0$. }
    \label{fig:Stoch_Time_Exotic_Phase_Functions}
\end{figure}

The different phases are classified as follows. A point $(m, \gamma, \delta)$ is said to be in Phase O if the supremum of eq. \eqref{eq:Stoch_Time_Work_Var_Nondim} 
on $s > 0$ 
is attained for some finite value of $s > 0$. That is if there is some $s_{\text{sup}} > 0$ such that $\mathcal{V}(s_{\text{sup}}) = \sup_{s  >0}\mathcal{V}(s)$. The point $(m, \gamma, \delta)$ is instead said to be in Phase B if it is not in Phase O and if the supremum is approached as $s \to 0^+$ {\it i.e.} $\lim_{s \to 0^+}\mathcal{V}(s) = \sup_{s  >0}\mathcal{V}(s)$. Lastly the point $(m, \gamma, \delta)$ is said to be in Phase G if it is neither in Phase O or Phase B. In this case, the supremum is approached uniquely as $s \to \infty$ or $\lim_{s \to \infty}\mathcal{V}(s) = \sup_{s  >0}\mathcal{V}(s)$. See Figures \ref{fig:StochTimePhaseFunction1}, \ref{fig:StochTimePhaseFunction2} and \ref{fig:Stoch_Time_Exotic_Phase_Functions} for different illustrations of how $\sigma_W^2$ vs $s$ could look like in the different phases.

\subsubsection{Phases Finite N}
We start the phase analysis by focusing on the case when $N$ is finite. It is also easier to divide into cases depending on whether or not $m = 1$.

\textbf{If} $m = 1$, we can rearrange eq. \eqref{eq:Stoch_Time_Work_Var_Nondim} into $\frac{\mathcal{V}(s)}{N^2}  =\frac{8+s\big(\gamma+N(\delta -1)-\delta\big)}{N(s+4)}$. By setting $K =\gamma+N(\delta -1)-\delta$ we find $\mathcal{V}'(s) = \frac{4K-8}{N(s+4)^2}$. Therefore when $K > 2$ we have that $\mathcal{V}'(s) > 0$ for all $s > 0$, so $\mathcal{V}(s)$ is strictly increasing. The supremum is then approached as $s \to \infty$. When $K = 2$ we find that $\mathcal{V}(s) = \frac{2}{N}$, which is a constant, this can be thought of as the supremum being attained for finite $s >0$ ( {\it e.g.} $s = 1$). If $K < 2$ we have that $\mathcal{V}'(s) < 0$ for all $s > 0$, implying that $\mathcal{V}(s)$ is strictly decreasing. The supremum is then approached as $s \to 0^+$.

Summarizing, when $m = 1$, Phase O is attained if and only if 
\begin{align}
    \label{eq:Stoch_Time_Phase_1_m_eq_1}
    \gamma+N(\delta -1)-\delta = 2.
\end{align}
Phase B is attained if and only if
\begin{align}
    \label{eq:Stoch_Time_Phase_2A_m_eq_1}
    \gamma+N(\delta -1)-\delta < 2.
\end{align}
Lastly, Phase G is attained if and only if
\begin{align}
    \label{eq:Stoch_Time_Phase_2B_m_eq_1}
    \gamma+N(\delta -1)-\delta > 2
\end{align}
This completes the analysis for $m = 1$.

\textbf{If} $m \neq 1$, we can divide both sides in eq. \eqref{eq:Stoch_Time_Work_Var_Nondim} by $N^2(m-1)^2$. We obtain:
\begin{align}
    \label{eq:Total_Work_Var_Setting_Nondim}
    \tilde{\mathcal{V}}(s)  
    &= \frac{2}{N(m-1)^2}+\frac{s}{s+2}B_N+\frac{s}{s+4}A_N-\frac{s^2}{(s+2)^2}, \\
    \label{eq:AN_Stochastic_Time}
    A_N &= \frac{\gamma-\delta -4m+2}{N(m-1)^2}+\frac{\delta-2m+1}{(m-1)^2}, \\
    \label{eq:BN_Stochastic_Time}
    B_N &=\frac{4}{N(m-1)}
\end{align}
We focus initially on obtaining the conditions for Phase O (in which the supremum is attained for finite $s > 0$). We first note that 
\begin{align}
    \label{eq:Work_Var_Stoch_Time_s_Infinite}
    \lim_{s \to \infty} \tilde{\mathcal{V}}(s)   = A_N+B_N-1+\frac{2}{N(m-1)^2}.
\end{align}
Next, we subtract eq. \eqref{eq:Work_Var_Stoch_Time_s_Infinite} from eq. \eqref{eq:Total_Work_Var_Setting_Nondim}, finding
\begin{align}
    \label{eq:Tot_Work_Var_Get_Phase_Trans}
    \tilde{\mathcal{V}}(s)-\lim_{s \to \infty} \tilde{\mathcal{V}}(s) &= \frac{4}{s+4}\big(\psi(s)-A_N\big) \text{ where } \\
    \notag 
    \psi(s) &= \frac{s+4}{4}\bigg(1-\frac{s^2}{(s+2)^2}-\frac{2B_N}{s+2} \bigg).
\end{align}
With eq. \eqref{eq:Tot_Work_Var_Get_Phase_Trans}, we will analyze the conditions that $A_N, B_N$ need to satisfy in order for the supremum to be attained for some finite $s > 0$. To this end, we need to compute $\sup_{s' > 0}\psi(s')$. 

Suppose the supremum of $\psi(s')$ is located on $(0, \infty)$. Then $\psi(s')$ should have a global maximum on $(0,\infty)$ at which the derivative is $0$. To find this global maximum, we differentiate $\psi(s')$, finding $\psi'(s') = \frac{2+2B_N-s'+B_Ns'}{(2+s')^3}$. The solution to $\psi'(s_c) = 0$ is therefore uniquely given by $s_c = \frac{2+2B_N}{1-B_N}$. This value of $s_c$ gives $\psi(s_c) = \frac{(-3+B_N)^2}{8}$ as a possible contender for the supremum of $\psi(s')$. If $|B_N| < 1$ then $\psi(s_c)$ will be the supremum. The reason is that then the unique stationary point $s_c$ satisfies $s_c \in (0,\infty)$, $\psi(s_c) \geq \lim_{s' \to \infty}\psi(s') = 1-\frac{B_N}{2}$ and $\psi(s_c) \geq \lim_{s' \to 0^+}\psi(s') =1-B_N$. If instead $|B_N| \geq 1$ it follows that $s_c$ is not finite or does not exist on $(0, \infty)$. In that case, the supremum can only be attained either at the start $s' = 0$ or at the end when $s' \to \infty$. We see that $\lim_{s' \to 0^+}\psi(s) = 1-B_N$ while $\lim_{s \to \infty} \psi(s) = 1-\frac{B_N}{2}$. Hence if $B_N \leq -1$ the supremum is expected at $s' = 0$ while for $B_N \geq 1$ the supremum is expected at $s' = \infty$. Summarizing, we have found
\begin{align}
    \label{eq:Supremum_Psi}
    \sup_{s' > 0} \psi(s') = 
    \begin{cases}
    1-B_N, \; &B_N \leq -1, \\
    \frac{(B_N-3)^2}{8}, &|B_N| < 1, \\
    1-\frac{B_N}{2}, &B_N \geq 1
    \end{cases}
\end{align}
With the aid of equations \eqref{eq:Tot_Work_Var_Get_Phase_Trans} and \eqref{eq:Supremum_Psi}, we can check if $\tilde{\mathcal{V}}(s)$ attains its supremum for finite $s > 0$ by dividing into three cases.
\\\\
(A) If $A_N > \sup_{s > 0} \psi(s)$ the supremum cannot ever be attained for finite $s > 0$. The reason is that in this case the expression in eq. \eqref{eq:Tot_Work_Var_Get_Phase_Trans} is less than zero for all $s > 0$ meaning that the limit $s \to \infty$ must yield the supremum. But that supremum obviously cannot be attained for finite $s > 0$.
\\\\
(B) If $A_N = \sup_{s > 0} \psi(s)$ we need to check if there is an interior point for which the supremum of $\psi(s')$ is attained. From the discussion before Eq. \eqref{eq:Supremum_Psi} we know that is the case when $|B_N| < 1$. So in the case that $A_N = \sup_{s > 0} \psi(s)$ and $|B_N| < 1$ the supremum is attained for finite $s > 0$ while if $A_N = \sup_{s > 0} \psi(s)$ and $|B_N| \geq 1$ that will not be the case. 
\\\\
(C) If instead $A_N <\sup_{s > 0} \psi(s)$ we have to be cautious. It could be that the supremum is not attained for any finite $s > 0$ because it is approached as $s \to 0^+$. To be certain that the supremum is attained for finite $s > 0$, we use Eq. \eqref{eq:Tot_Work_Var_Get_Phase_Trans} to compute
\begin{align}
    \label{eq:Tot_Work_Var_Get_Phase_Trans_Check_0}
    \tilde{\mathcal{V}}(s)-\lim_{s \to 0^+}\tilde{\mathcal{V}}(s) = \frac{s\big((A_N+B_N-1)s^2+(4A_N+6B_N-4)s+4(A_N+2B_N)\big)}{(s+4)(s+2)^2}.
\end{align}
The question of whether the supremum is attained for $s > 0$ in the regime $A_N <\sup_{s > 0} \psi(s)$ can be divided into three cases depending on the sign of $A_N+B_N-1$.

(i) When $A_N+B_N-1 > 0$ the supremum  cannot be attained at $s \to 0^+$. To see this, first note that in this case $\lim_{s \to \infty}  \tilde{\mathcal{V}}(s)-\lim_{s \to 0^+}\tilde{\mathcal{V}}(s)= A_N+B_N-1 > 0$. So the supremum of $\tilde{\mathcal{V}}(s)$ cannot be approached as $s \to 0^+$ since that limiting value was less than the value of $\tilde{\mathcal{V}}(s)$ as $s \to \infty$. But in the regime $A_N < \sup_{s > 0} \psi(s)$, the supremum was not attained as $s \to \infty$. In conclusion, the supremum must be attained for some finite $s > 0$. 

(ii) When $A_N+B_N-1 = 0$ we see that eq. \eqref{eq:Tot_Work_Var_Get_Phase_Trans_Check_0} gives $\tilde{\mathcal{V}}(s)-\lim_{s \to 0^+}\tilde{\mathcal{V}}(s) = \frac{s\big(2B_Ns+4(B_N+1)\big)}{(s+4)(s+2)^2}$. This expression tells us that $s \to 0^+$ gives the supremum only if $B_N \leq -1$. In other cases, the supremum will be attained for finite $s > 0$.

(iii) In the regime $A_N+B_N-1 < 0$ we see that $\tilde{\mathcal{V}}(s)-\frac{2}{N(m-1)^2} \leq 0$ for all $s > 0$ will hold if and only if $A_N+2B_N \leq 0$. The reason is that the quadratic term in \eqref{eq:Tot_Work_Var_Get_Phase_Trans_Check_0} is always negative due to $A_N+B_N-1 < 0$ and in the case $A_N+2B_N \leq 0$ the constant term is non-positive. Note that
\begin{align*}
    4A_N+6B_N-4 = 2(A_N+B_N-1)+2(A_N+2B_N)-2<0    
\end{align*}
whenever $A_N+2B_N \leq 0$. So this shows that we do expect $s \to 0^+$ to be the supremum when $A_N+2B_N \leq 0$. If instead $A_N+2B_N > 0$ then by choosing $s$ very close to $0$, we can make $(A_N+B_N-1)s^2+(4A_N+6B_N-4)s+4(A_N+2B_N) > 0$ thus implying that $s \to 0^+$ will not give the supremum. 

Combining (A), (B) and (C) together, we have found that the supremum is attained for finite $s > 0$ if and only if either of these cases hold
\begin{align}
    \begin{cases}
       A_N = \sup_{s > 0} \psi(s) & |B_N| < 1 \\
       A_N <\sup_{s > 0} \psi(s) & A_N+B_N-1 > 0 \\
       A_N <\sup_{s > 0} \psi(s) & A_N+B_N-1 = 0, B_N > -1 \\
       A_N  <\sup_{s > 0} \psi(s) & A_N+B_N-1 < 0, A_N+2B_N > 0
    \end{cases}
\end{align}
Substituting eq. \eqref{eq:Supremum_Psi} into here, we find at first the following conditions:
\begin{align*}
    \begin{cases}
       A_N = \frac{(B_N-3)^2}{8}, & |B_N| < 1 \\
       1-B_N< A_N < 1-B_N, & B_N \leq -1 \\
       1-B_N< A_N < \frac{(B_N-3)^2}{8}, &|B_N| < 1 \\
       1-B_N< A_N < 1-\frac{B_N}{2}, &B_N \geq 1\\
       1-B_N= A_N < \frac{(B_N-3)^2}{8} &  |B_N| < 1\\
       1-B_N= A_N <1-\frac{B_N}{2}& B_N \geq 1 \\
       -2B_N< A_N  < 1-B_N  &  B_N \leq -1, \;A_N+B_N-1 < 0 \\
       -2B_N<A_N  < \frac{(B_N-3)^2}{8} & |B_N| < 1, \;A_N+B_N-1 < 0\\
       -2B_N<A_N  <  1-\frac{B_N}{2} &  B_N \geq 1, \;A_N+B_N-1 < 0
    \end{cases}
\end{align*}
We can however greatly simplify and reduce the number of above conditions. First note that the cases that have $B_N \leq - 1$ included as a condition are all impossible. Second, we can combine all cases with $B_N \geq 1$ since in that regime $1-B_N > -2B_N$ and $1-\frac{B_N}{2} \geq 1- B_N$. Summarizing, we get
\begin{align*}
    \begin{cases}
       A_N = \frac{(B_N-3)^2}{8}, & |B_N| < 1 \\
       1-B_N< A_N < \frac{(B_N-3)^2}{8}, &|B_N| < 1 \\
       1-B_N= A_N < \frac{(B_N-3)^2}{8} &  |B_N| < 1\\
       -2B_N<A_N  < \frac{(B_N-3)^2}{8} & |B_N| < 1, \;A_N+B_N-1 < 0\\
       -2B_N<A_N  <  1-\frac{B_N}{2} &  B_N \geq 1
    \end{cases}
\end{align*}
We proceed by combining some cases with $|B_N| < 1$. The reason is that in this regime, $1-B_N > -2B_N$ and also $\frac{(B_N-3)^2}{8} > -2B_N$. We find that
\begin{align}
    \label{eq:App_Stoch_Time_Phase_1_m_not_1}
    \begin{cases}
       -2B_N<A_N  \leq \frac{(B_N-3)^2}{8} & |B_N| < 1\\
       -2B_N<A_N  <  1-\frac{B_N}{2} &  B_N \geq 1 
    \end{cases}
\end{align}
Hence, when either of the conditions in eq. \eqref{eq:App_Stoch_Time_Phase_1_m_not_1} is satisfied for $A_N = \frac{\gamma-\delta -4m+2}{N(m-1)^2}+\frac{\delta-2m+1}{(m-1)^2}$ and $B_N = \frac{4}{N(m-1)}$, then we know that the supremum is attained for finite $s > 0$. Outside this region, the supremum is not attained for finite $s > 0$ but only in the limits $s \to 0^+$ or $s \to \infty$. Thus the cases in eq. \eqref{eq:App_Stoch_Time_Phase_1_m_not_1} define Phase O.

\textbf{Phase O}: Combining the cases for Phase O when $m = 1$, stated in eq. \eqref{eq:Stoch_Time_Phase_1_m_eq_1}, together with the cases for Phase O when $m \neq 1$, stated in eq. \eqref{eq:App_Stoch_Time_Phase_1_m_not_1} we find that a point $(m, \gamma, \delta)$ is in Phase O if and only if
  \begin{align}
        \label{eq:App_Stoch_Time_Phase_1}
        \begin{cases}
            \gamma +(N-1)\delta = N+2 &m=1, \\
            -2B_N<A_N\leq \frac{(B_N-3)^2}{8} &m\neq 1, \; -1 < B_N < 1 \\
            -2B_N<A_N<1-\frac{B_N}{2} &m\neq 1, \quad B_N \geq 1
        \end{cases}
    \end{align}
Outside the region specified by Eq. \eqref{eq:App_Stoch_Time_Phase_1} the supremum is approached as $s \to 0^+$ or $s \to \infty$. 

Let us now continue the analysis by investigating when $(m, \gamma, \delta)$ belongs to Phase B.
When $m = 1$ we see that Phase B is attained when eq. \eqref{eq:Stoch_Time_Phase_2A_m_eq_1} holds. Suppose therefore that $m \neq 1$. We deduce by looking at condition (C) above,  that when $m \neq 1$ the supremum is approached as $s \to 0^+$ if and only if 
\begin{align}
    \label{eq:App_Stoch_Time_Phase_2A_Finite_Intermediate}
    \begin{cases}
        A_N+B_N = 1, &m \neq1, \; B_N \leq -1 \\
        A_N+B_N <1 &m \neq 1, \; A_N+2B_N \leq 0
    \end{cases}
\end{align}
Observe that the condition $A_N+B_N = 1$ and $B_N \leq -1$ implies $A_N+2B_N \leq 0$ meaning that the two rows in eq. \eqref{eq:App_Stoch_Time_Phase_2A_Finite_Intermediate} can be combined into 
\begin{align}
    \label{eq:App_Stoch_Time_Phase_2A_Finite_Intermediate_2}
    A_N+B_N \leq 1 \,\,  &m \neq 1, \; A_N+2B_N \leq 0.
\end{align}
\textbf{Phase B}: Combining the cases for Phase B when $m = 1$, stated in eq. \eqref{eq:Stoch_Time_Phase_2A_m_eq_1}, together with the cases for Phase B when $m \neq 1$, stated in eq. \eqref{eq:App_Stoch_Time_Phase_2A_Finite_Intermediate_2}, we find that a point $(m, \gamma, \delta)$ is in Phase B if and only if
\begin{align}
    \label{eq:App_Stoch_Time_Phase_2A_Finite}
    \begin{cases}
        \gamma + N(\delta-1)-\delta <2&m = 1\\
        A_N+B_N \leq 1 &m \neq 1, \; A_N+2B_N \leq 0
    \end{cases}
\end{align}
We proceed to the last phase, namely when $(m, \gamma, \delta)$ belongs to Phase G.

When $m = 1$ we see that Phase G is attained when eq. \eqref{eq:Stoch_Time_Phase_2B_m_eq_1} holds.

Whenever $m \neq 1$ we need to check the conditions in (A) or (B). Doing that we see that the supremum is not attained for finite $s > 0$ if $A_N > \sup_{s > 0}\psi(s)$ or if $A_N = \sup_{s >0}\psi(s)$ together with $|B_N| \geq 1$ holds. Looking at eq. \eqref{eq:Tot_Work_Var_Get_Phase_Trans} we see that the supremum will be approached as $s \to \infty$ and not $s \to 0^+$ in this regime if $1-B_N-A_N < 0$. From eq. \eqref{eq:Supremum_Psi} we see that these conditions tell us that when $m\neq 1$ we expect Phase G if and only if
\begin{align}
    \begin{cases}
        A_N > 1-B_N & B_N \leq -1, \; A_N+B_N >1 \\
        A_N > \frac{(B_N-3)^2}{8} & |B_N| < 1, \; A_N+B_N >1 \\
        A_N > 1-\frac{B_N}{2}& B_N \geq 1, \; A_N+B_N >1 \\
        A_N = 1-B_N & B_N \leq -1, \; A_N+B_N >1\\
        A_N = 1-\frac{B_N}{2} & B_N \geq 1, \; A_N+B_N > 1
    \end{cases}
\end{align}
For the first row here, note that $A_N+B_N > 1$ is already enforced by $A_N > 1-B_N$ so it can be dropped. 

In the second row, we check whether $\frac{(B_N-3)^2}{8}$ or $1-B_N$ gives the greatest lower bound on $A_N$. It turns out that $\frac{(B_N-3)^2}{8}-(1-B_N) = \frac{B_N^2+2B_N+1}{8} \geq 0$ so $\frac{(B_N-3)^2}{8}$ always gives the greatest lower bound so we can neglect $A_N+B_N > 1$ in this row.

For the third row, we see that whenever $A_N > 1-\frac{B_N}{2}$ and $B_N \geq 1$ it follows that $A_N+B_N > 1+\frac{B_N}{2} \geq \frac{3}{2} > 1$ so we can drop the condition $A_N+B_N > 1$.

In the fourth row here, we have on the one hand $A_N = 1-B_N$ on the other $A_N > 1-B_N$ which is a contradiction. 

For the fifth row here, we have from $A_N = 1-\frac{B_N}{2}$ and $B_N \geq 1$ that $A_N+B_N = 1+\frac{B_N}{2} \geq \frac{3}{2} > 1$. So we can on the fifth row drop the condition $A_N+B_N >1$. We can then combine the fifth row with the third row. 

In summary, we find
\begin{align}
    \label{eq:Stoch_Time_Phase_2B_m_meq_1}
    \begin{cases}
        A_N > 1-B_N & B_N \leq -1 \\
        A_N > \frac{(B_N-3)^2}{8} & |B_N| < 1 \\
        A_N \geq 1-\frac{B_N}{2}& B_N \geq 1 \\
    \end{cases}
\end{align}
\textbf{Phase G}: Combining the conditions for Phase G when $m = 1$, stated in eq. \eqref{eq:Stoch_Time_Phase_2B_m_eq_1}, with the conditions for Phase G when $m \neq 1$, stated in eq. \eqref{eq:Stoch_Time_Phase_2B_m_meq_1}, we find that a point $(m, \gamma, \delta)$ belongs to Phase G if and only if
\begin{align}
    \begin{cases}
        \label{eq:App_Stoch_Time_Phase_2B_Finite}
        \gamma + N(\delta-1)-\delta >2 & m=1 \\
        A_N > 1-B_N & m\neq 1, \; B_N \leq -1 \\
        A_N > \frac{(B_N-3)^2}{8} & m\neq 1, \;  |B_N| < 1 \\
        A_N \geq 1-\frac{B_N}{2}& m\neq 1, \;  B_N \geq 1
    \end{cases}
\end{align}
We have therefore completed the phase analysis for finite $N$. Summarizing, the point $(m, \gamma, \delta)$ belongs to Phase O if and only if the conditions in eq. \eqref{eq:App_Stoch_Time_Phase_1} hold. It instead belongs to Phase B if and only if the conditions in eq. \eqref{eq:App_Stoch_Time_Phase_2A_Finite} hold. Lastly, it belongs to Phase G if and only if the conditions in eq. \eqref{eq:App_Stoch_Time_Phase_2B_Finite} holds.
\subsubsection{Phases Infinite N}
When we take $N \to \infty$ in Eq. \eqref{eq:App_Stoch_Time_Phase_1}, we have to be cautious since depending on the values of $m,  \gamma, \delta$ the expression scales differently with $N$.  To take the limits appropriately, we need to hold $m$, $\gamma$ and $\delta$ fixed in the beginning and then take $N \to \infty$. 

We begin by taking the $N \to \infty$ limit of the predicates of Phase O in Eq. \eqref{eq:App_Stoch_Time_Phase_1}. We also again separate the cases $m = 1$ and $m \neq 1$.

\textbf{If}: $m = 1$, eq. \eqref{eq:App_Stoch_Time_Phase_1} gives the condition
\begin{align*}
    \gamma +(N-1)\delta = N+2.
\end{align*}
Now if $\delta \neq 1$ and if $m, \gamma, \delta$ are held fixed this equation cannot hold as $N \to \infty$.

If instead $\delta = 1$ then this equation gives us immediately $\gamma = 3$.

So when $m = 1$ we know that the point $(m = 1, \gamma = 3, \delta = 1)$ will belong to Phase O. All other points here, $(m = 1, \gamma, \delta)$ with $\gamma \neq 3$ or $\delta\neq 1$ will belong to Phase B or Phase G so no supremum is attained for finite $s > 0$.

\textbf{If}: $m \neq 1$ we know in the $N \to \infty$ limit that $B_N \to 0$ independently of $\gamma, \delta$ by Eq. \eqref{eq:BN_Stochastic_Time}. So in this case, the condition $-1 < B_N < 1$ in Eq. \eqref{eq:App_Stoch_Time_Phase_1} will hold eventually as $N \to \infty$ for any $(m, \gamma, \delta)$. Thus in the $N \to \infty$ limit we have that $(m, \gamma, \delta)$ is in Phase O if and only if $-2B_N < A_N\leq \frac{(B_N-3)^2}{8}$. Substituting $A_N = \frac{\gamma-\delta -4m+2}{N(m-1)^2}+\frac{\delta-2m+1}{(m-1)^2}$ and $B_N = \frac{4}{N(m-1)}$ into this inequality, we find Phase O when
\begin{align*}
    -\frac{8}{N(m-1)} < \frac{\gamma-\delta -4m+2}{N(m-1)^2}+\frac{\delta-2m+1}{(m-1)^2} \leq \frac{( \frac{4}{N(m-1)}-3)^2}{8}.
\end{align*}
Simplifying this inequality by multiplying with $(m-1)^2$ on both sides, we obtain
\begin{align}
   \label{eq:App_Stoch_Time_Phase_1_Infinite_Intermediate}
   -\frac{8}{N}(m-1) < \frac{\gamma -\delta -4m+2}{N}+\delta-2m+1 \leq \frac{2}{N^2}-\frac{3}{N}(m-1)+\frac{9}{8}(m-1)^2.
\end{align}
If $\delta -2m+1 \neq \frac{9}{8}(m-1)^2$ then the limit $N \to \infty$ of Eq. \eqref{eq:App_Stoch_Time_Phase_1_Infinite_Intermediate} will give 
\begin{align}
    \label{eq:App_Stoch_Time_Phase_1_Infinite_Intermediate_Part_1}
    \delta -2m+1 < \frac{9}{8}(m-1)^2 \;.
\end{align}

The case $\delta -2m+1 = \frac{9}{8}(m-1)^2$ is more difficult. If this occurs, Eq. \eqref{eq:App_Stoch_Time_Phase_1_Infinite_Intermediate} becomes
\begin{align*}
    -8(m-1)-\frac{9}{8}(m-1)^2N<\gamma -\delta -4m+2\leq \frac{2}{N}-3(m-1).
\end{align*}
Now, we can take $N \to \infty$. We find that the lower bound becomes $-\infty$ so we can ignore it, finding
\begin{align*}
    \gamma -\delta -4m+2\leq -3(m-1).
\end{align*}
Using $\delta -2m+1 = \frac{9}{8}(m-1)^2$ we find here
\begin{align*}
    \gamma -6m+3 \leq \delta -5m+4.
\end{align*}
This inequality implies that
\begin{align}
    \label{eq:App_Stoch_Time_Phase_1_Infinite_Intermediate_Part_2}
    \gamma \leq \delta +m+1.
\end{align}
With Eq. \eqref{eq:App_Stoch_Time_Phase_1_Infinite_Intermediate_Part_1} when $\delta-2m+1 \neq \frac{9}{8}(m-1)^2$ and eq. \eqref{eq:App_Stoch_Time_Phase_1_Infinite_Intermediate_Part_2} when $\delta-2m+1 = \frac{9}{8}(m-1)^2$, we have fully dealt with the $m \neq 1$ case.

\textbf{Phase O}: Summarizing, for the $N \to \infty$ case, we have found that a point $(m,\gamma, \delta)$ is in Phase O if and only if
\begin{align}
    \label{eq:App_Stoch_Time_Phase_1_Infinite_N}
    \begin{cases}
        \delta -2m+1 < \frac{9}{8}(m-1)^2, \; &m\neq 1 \\
         \gamma \leq \delta +m+1 &m\neq 1, \; \delta -2m+1 = \frac{9}{8}(m-1)^2 \\
         \gamma = 3, \delta = 1 &m=1.
    \end{cases}
\end{align}
For Phase B, we take $N \to \infty$ of eq. \eqref{eq:App_Stoch_Time_Phase_2A_Finite}. \textbf{If}: $m = 1$ we find that $\delta = 1$ becomes enforced and also $\gamma < 3$.

\textbf{If}: $m \neq 1$ we need to check how $A_N+B_N$ and $A_N+2B_N$ appear in the limit $N \to \infty$. We get for example that
\begin{align*}
    A_N+2B_N \to \begin{cases}
        \frac{\delta -2m+1}{(m-1)^2} & \text{if } \delta -2m+1 \neq 0 \\
        \frac{\gamma +2m-5}{N(m-1)^2} & \text{if } \delta-2m+1 = 0
    \end{cases}
\end{align*}
as $N \to \infty$. So to have $A_N+2B_N \leq 0$ as $N \to \infty$ we require $\delta-2m+1 \leq 0$. When $\delta-2m+1 = 0$ we also require $\gamma +2m-5 \leq 0$. Otherwise, we would find that $A_N+2B_N = \frac{\gamma+2m-5}{N(m-1)^2} > 0$ for all $N$ meaning that we cannot ever satisfy $A_N+B_N \leq 0$ for large enough $N$.

For the other case or $A_N+B_N$ we find
\begin{align*}
    A_N+B_N \to \begin{cases}
        \frac{\delta -2m+1}{(m-1)^2} & \text{if } \delta -2m+1 \neq 0 \\
        \frac{\gamma - \delta -4m+2+4(m-1)}{N(m-1)^2} & \text{if } \delta-2m+1 = 0
    \end{cases}
\end{align*}
as $N \to \infty$. Note that this means that $A_N+B_N \leq 1$ is satisfied as $N \to \infty$ if $\delta -2m+1 \leq 0$. We therefore just have to focus on $A_N+2B_N \leq 0$.

So we have Phase B if and only if
\begin{align}
    \label{eq:App_Stoch_Time_Phase_2A_Infinite_Intermediate}
    \begin{cases}
        \gamma < 3,\;\delta = 1 & m= 1, \\
        \delta -2m+1<0& m\neq 1 \\
        \gamma +2m-5 \leq 0,\;  \delta -2m+1 = 0 & m\neq 1
    \end{cases}
\end{align}
The last two rows in Eq. \eqref{eq:App_Stoch_Time_Phase_2A_Infinite_Intermediate} are however impossible to satisfy. The reason is that on the variables $(m, \gamma, \delta)$ we have the condition $\gamma -m+(N-1)(\delta -m^2) \geq 0$. Therefore as $N \to \infty$ we require $\delta \geq m^2$. But this means that $\delta -2m+1 = \delta -m^2 +(m-1)^2 > 0$ whenever $m \neq 1$, ensuring that the last two rows in eq. \eqref{eq:App_Stoch_Time_Phase_2A_Infinite_Intermediate} are impossible. 

\textbf{Phase B} thus appears if and only if 
\begin{align}
    \label{eq:App_Stoch_Time_Phase_2A_Infinite}
    \gamma < 3,\;\delta = 1, \; m= 1. 
\end{align}
For Phase G, we want to take the $N\to \infty$ limit of Eq. \eqref{eq:App_Stoch_Time_Phase_2B_Finite}. 

\textbf{If}: $m = 1$ we see that the first row forces us to have $\delta \geq 1$. If $\delta = 1$ the first row of eq. \eqref{eq:App_Stoch_Time_Phase_2B_Finite} becomes $\gamma> 3$. If $\delta > 1$ we find that $\gamma$ take on any of its admissible values.

\textbf{If}: $m \neq 1$ we know that $B_N \to 0$ independently of what $m, \gamma, \delta$ by eq. \eqref{eq:BN_Stochastic_Time}. So this means that only the condition $|B_N| < 1$ can be satisfied as $N \to \infty$. We therefore need to check what happens to $A_N > \frac{(B_N-3)^2}{8} $ as $N \to \infty$. Substituting in $A_N = \frac{\gamma-\delta -4m+2}{N(m-1)^2}+\frac{\delta-2m+1}{(m-1)^2}$ and $B_N = \frac{4}{N(m-1)}$ we find
\begin{align*}
     \frac{\gamma-\delta -4m+2}{N(m-1)^2}+\frac{\delta-2m+1}{(m-1)^2}>\frac{1}{8}\Big(9-\frac{12}{N(m-1)}+\frac{16}{N^2(m-1)^2} \Big).
\end{align*}
If $\frac{\delta-2m+1}{(m-1)^2} \neq \frac{9}{8}$ as we take $N \to \infty$ this condition tells us that
\begin{align*}
    \frac{\delta-2m+1}{(m-1)^2} = \frac{9}{8}.
\end{align*}
If instead $\frac{\delta-2m+1}{(m-1)^2} = \frac{9}{8}$ we find as $N \to \infty$ the condition below should hold:
\begin{align*}
    \gamma-\delta -m-1  >0.
\end{align*}
\textbf{Phase G}: Combining everything, Phase G is attained if and only if
\begin{align}
    \begin{cases}
        \label{eq:App_Stoch_Time_Phase_2B_Infinite}
        \delta > 1 & m= 1\\
        \gamma > 3& m = 1, \; \delta =1 \\
        \frac{\delta-2m+1}{(m-1)^2} > \frac{9}{8}& m\neq 1
        \\
        \gamma - \delta -m > 1, & m\neq 1, \; \frac{\delta-2m+1}{(m-1)^2} = \frac{9}{8}
    \end{cases}.
\end{align}
\subsection{Stochastic spring constant}
Using eq. \eqref{eq:Work_Variance_Stoch_kV}, we find for the non-dimensionalized work variance when using the abbreviations $\xi = \braket{(k_V-k_U)^2}$ and $\chi = \braket{k_V-k_U}^2$ the expression
\begin{align}
    \notag
    \mathcal{V} &= \xi\Bigg(Nme^{-2\tau^* }u(\tau^*)+\frac{1}{4}\Big(N\gamma+N(N-1)\delta \Big) e^{-4\tau^*}+\frac{1}{4}(N^2+2N)u(\tau^*)^2+ \\
    \label{eq:Non-Dim_Work_Var_Stoch_kV}
    &+\frac{1}{2}N^2me^{-2\tau^*}u(\tau^*) \Bigg)  
    -\frac{1}{4}N^2\chi\Bigg(u(\tau^*)+me^{-2\tau^*} \Bigg)^2.
\end{align}
Eq. \eqref{eq:Non-Dim_Work_Var_Stoch_kV}  will be utilized to analyze the phases. This is just Eq. (9) of the main article. Reminder that here $u(\tau^*) =1-e^{-2\tau^*}$.

\begin{figure}[htbp]
    \centering
    \begin{minipage}[b]{0.35\textwidth}
        \centering
        \includegraphics[width=\linewidth]{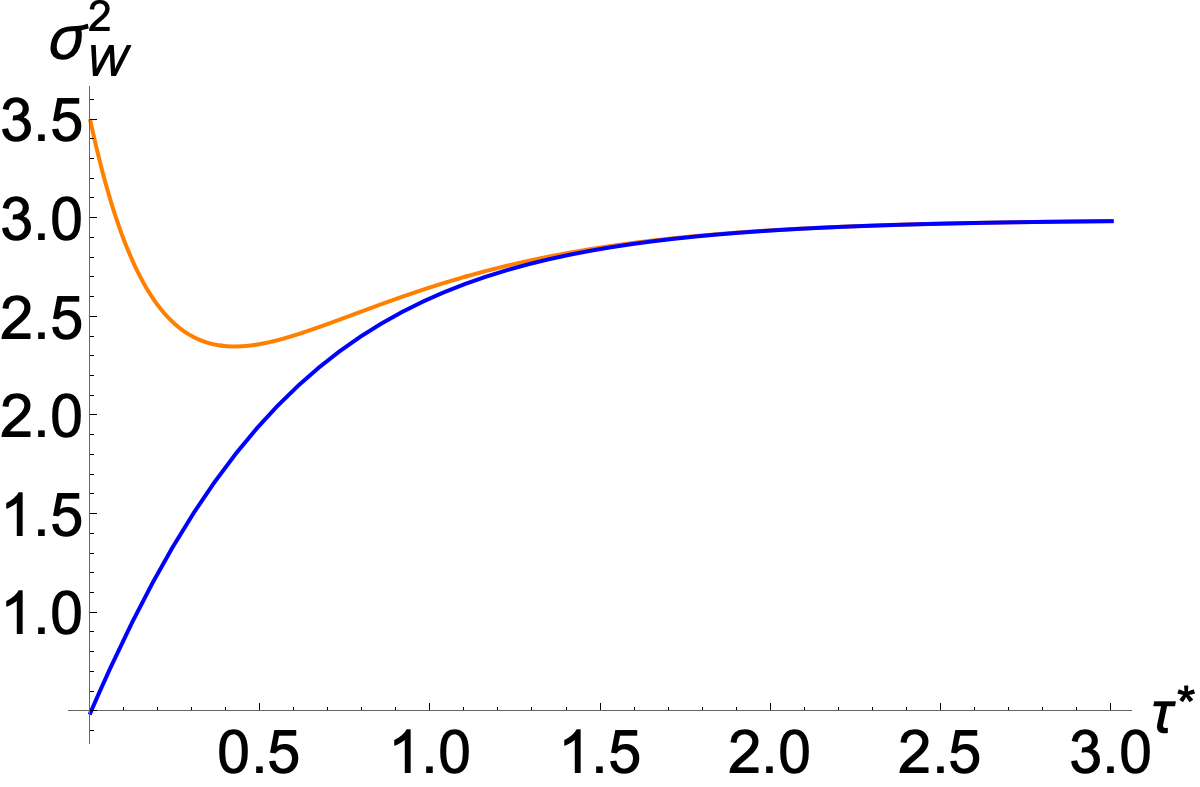}
        \caption{This shows an example of the form of $\mathcal{V}(s)$ or Eq. \eqref{eq:Non-Dim_Work_Var_Stoch_kV} for $(m, \gamma, \delta) = (\frac{1}{2},\frac{1}{2},\frac{1}{4})$ in Phase B and also an example for $(m, \gamma, \delta) = (\frac{1}{2},\frac{7}{2},\frac{1}{4})$ in Phase O. In both cases, we have $N = 2$, $\xi = 2$ and $\chi = 1$.}
        \label{fig:StochkVPhaseFunction1}
    \end{minipage}
    \hspace{1cm}
    \begin{minipage}[b]{0.35\textwidth}
        \centering
        \includegraphics[width=\linewidth]{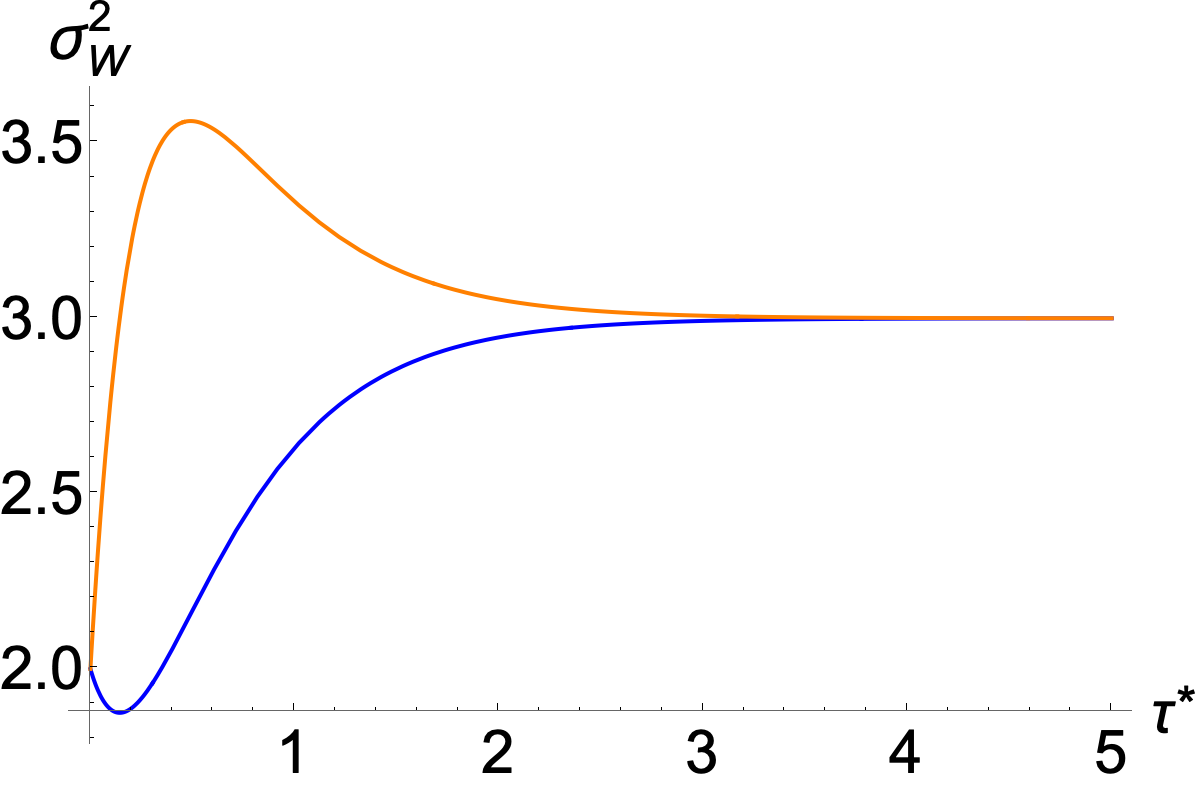}
        \caption{This shows an example of the form of $\mathcal{V}(s)$ or Eq. \eqref{eq:Non-Dim_Work_Var_Stoch_kV} for $(m, \gamma, \delta) = (\frac{3}{2},2,\frac{9}{4})$ in Phase B and also an example for $(m, \gamma, \delta) = (\frac{1}{2},2,\frac{1}{4})$ in Phase O. In both cases, we have $N = 2$, $\xi = 2$ and $\chi = 1$.}
        \label{fig:StochkVPhaseFunction2}
    \end{minipage}
\end{figure}

The different phases are classified as follows. A point $(m, \gamma, \delta)$ is said to be in Phase O if the supremum of Eq. \eqref{eq:Stoch_Time_Work_Var_Nondim} on $\tau^* \geq 0$ is attained for some finite value of $\tau^* \geq 0$. That is if there is some $\tau^*_{\text{sup}} \geq 0$ such that $\mathcal{V}(\tau^*_{\text{sup}}) = \sup_{\tau^*  \geq0}\mathcal{V}(\tau^*)$. The point $(m, \gamma, \delta)$ is instead said to be in Phase B if it is not in Phase O and the supremum is attained as $\tau^* \to \infty$. Or $\lim_{\tau^* \to \infty}\mathcal{V}(\tau^*) = \sup_{\tau^*  \geq 0}\mathcal{V}(\tau^*)$. For an illustration of these phases, see Figures \ref{fig:StochkVPhaseFunction1} and \ref{fig:StochkVPhaseFunction2}.

We start with restricting ourselves to finite $N$.
\subsubsection{Phases Finite N}
We see that in Eq. \eqref{eq:Non-Dim_Work_Var_Stoch_kV}, we have three dynamical parameters: $\braket{k_V^2}$, $\braket{k_V}$ (through both $\xi$ and $\chi$) and $\tau^*$ while we have four parameters from the initial conditions: $N$, $m$, $\gamma$ and $\delta$. To analyze the phases, we have to check whether some values of the initial conditions ensure that no matter the value of the dynamical parameters, the supremum is either not attained for finite values of these dynamical parameters or it is.

The factor in front of $\xi$ in eq. \eqref{eq:Non-Dim_Work_Var_Stoch_kV} is positive except when $\tau^* = 0$, $\gamma = 0$ and $\delta  =0$. Since $\gamma \sim \braket{x_{01}^4}$ the only possibility here is that $P(x_{01})=\delta(x_{01})$. Therefore, $\sigma_W^2 = 0$ follows no matter what $\braket{k_V^2}$ and $\braket{k_V}$ are. What this establishes is that if we are allowed to vary $\braket{k_V^2}$ independently, Eq. \eqref{eq:Non-Dim_Work_Var_Stoch_kV} can be made arbitrarily large. This means that the phase structure is not interesting if we allow to vary $\braket{k_V^2}$ independently, almost every point $(m, \gamma, \delta)$ (except the special point where $\sigma_W^2$ is constant) belongs to the phase where the supremum is approached as $\braket{k_V^2} \to \infty$. This is all unless we are on the condition $\tau^* = 0$, $m = \gamma = \delta = 0$. Then the supremum is attained for all values of $\braket{k_V^2}$ and $\braket{k_V}$.

It is more interesting to analyze the case when we fix the values of $\braket{k_V^2}$ and $\braket{k_V}$ and allow only $\tau^*$ to change. We see first that Eq. \eqref{eq:Non-Dim_Work_Var_Stoch_kV} can be rewritten as
\begin{align}
    \notag
    \mathcal{V} &= \frac{N}{4}\big((N+2)\xi-N\chi\big)+\frac{N}{2}e^{-2\tau^*}(m-1)\big((N+2)\xi-N\chi \big) + \\
    &+\frac{N}{4}e^{-4\tau^*}\Big(\big(2+N-2m(2+N)+\gamma +(N-1)\delta\big)\xi-(m-1)^2N\chi\Big).
\end{align}
To investigate whether we have different phases in this expression, we introduce $z = e^{-2\tau^*}$ finding the quadratic
\begin{align}
    \notag
    \mathcal{V} &= \frac{N}{4}\big((N+2)\xi-N\chi\big)+\frac{N}{2}z(m-1)\big((N+2)\xi-N\chi \big) + \\
    \label{eq:Phase_Trans_Stoch_kV}
    &+\frac{N}{4}z^2\Big(\big(2+N-2m(2+N)+\gamma +(N-1)\delta\big)\xi-(m-1)^2N\chi\Big).
\end{align}
We will analyze when Eq. \eqref{eq:Phase_Trans_Stoch_kV} attains its supremum in $z \in (0, 1]$.

For brevity, let us set $\bar{A}_N = \frac{N}{4}\Big(\big(2+N-2m(2+N)+\gamma +(N-1)\delta\big)\xi-(m-1)^2N\chi\Big)$, $\bar{B}_N = \frac{N}{2}(m-1)\big((N+2)\xi-N\chi \big)$ and $\bar{C}_N = \frac{N}{4}\big((N+2)\xi-N\chi\big)$. Then this equation tells us that 
\begin{align}
    \label{eq:Quadratic_Phase_Trans_Stoch_kV}
    \tilde{\mathcal{V}}(z) = \bar{C}_N+\bar{B}_Nz+\bar{A}_Nz^2.    
\end{align}
If $\bar{B}_N  > 0$ then this function will attain its supremum on $(0,1]$. The reason is that we know that there should be some small $1 \geq z_1 > 0$ such that $\bar{B}_Nz_1 \geq -\bar{A}_Nz_1^2$. This implies that $\tilde{\mathcal{V}}$ cannot attain its supremum uniquely at $z = 0$ since then $\tilde{\mathcal{V}}(z_1) \geq \tilde{\mathcal{V}}(z =0)$. The function must  attain its supremum in some $1 \geq \tilde{z} > 0$ due to this, it could however attain its supremum in two points $0$ and $\tilde{z}$. 

If $\bar{B}_N \leq 0$ then the supremum will be attained on $(0,1]$ if and only if $\bar{A}_N + \bar{B}_N \geq 0$. To see this, note that the supremum will be attained for finite $\tau^*$ if $\tilde{\mathcal{V}}(0) \leq \tilde{\mathcal{V}}(\tilde{z})$ for some $1 \geq \tilde{z} > 0$. This provides us with the condition $\tilde{z}(\bar{B}_N+\bar{A}_N\tilde{z}) \geq 0$. Dividing both sides by  $\tilde{z}$ since $1 \geq \tilde{z} > 0$ we have $\bar{A}_N \tilde{z} \geq -\bar{B}_N$. Note that since $\bar{B}_N \leq 0$ we conclude that $\bar{A}_N \tilde{z}  \geq 0$ follows. Because $1 \geq \tilde{z} >0$ this means that $\bar{A}_N \geq -\bar{B}_N$. In fact, we can see that if $\bar{A}_N < -\bar{B}_N$ then $\bar{B}_N + \bar{A}_Nz < \bar{B}_N(1-z) \leq 0$ for all $1\geq z\geq0$. That would then imply $\tilde{\mathcal{V}}(z) = \bar{C}_N+\bar{B}_Nz + \bar{A}_Nz^2 < \bar{C}_N = \tilde{\mathcal{V}}(0)$ for all $1 \geq z > 0$ so the supremum is attained at the inaccessible point $z = 0$ (which corresponds to $\tau^* = -\infty$).
 
So summarizing, we have found that the supremum is attained for $1 \geq z > 0$ and consequently for finite $\tau^* \geq 0$ if and only if
\begin{align}
    \label{eq:Phase_Stoch_kV}
    \begin{cases}
        \bar{B}_N > 0, \\
        \bar{A}_N+\bar{B}_N \geq 0
    \end{cases}.
\end{align}
Substituting the equations for $\bar{A}_N$ and $\bar{B}_N$, we find
\begin{align}
    \label{eq:Phase_Stoch_kV_1}
    \begin{cases}
        m > 1, \\
        \big(-2+\gamma+N(\delta-1)-\delta\big)\xi -(m^2-1)N\chi \geq 0
    \end{cases}.
\end{align}
The first row in Eq. \eqref{eq:Phase_Stoch_kV_1} follows from the fact that $\bar{B}_N \propto (m-1)$ along with all other factors being not negative implies that $\bar{B}_N > 0$ holds if and only if $m > 1$.

It is possible to simplify the second condition in Eq. \eqref{eq:Phase_Stoch_kV_1}. First, introduce $C_N = \gamma-m^2+(N-1)(\delta -m^2)$. With that, Eq. \eqref{eq:Phase_Stoch_kV_1} reads
\begin{align*}
    \begin{cases}
        m > 1, \\
        \big(-2-N+Nm^2+C_N\big)\xi -(m^2-1)N\chi \geq 0
    \end{cases}.
\end{align*}
If we assume that $\xi > 0$ we can rearrange this expression finding
\begin{align*}
    \begin{cases}
        m > 1, \\
       C_N \geq 2+\frac{N(1-m^2)}{\xi}(\xi-\chi)
    \end{cases}.
\end{align*}
Note that it is no problem assuming $\xi > 0$ since if $\xi = 0$ we would have $\xi = \braket{(k_V-k_U)^2} = 0$ 
implying $k_V =k_U$.
We do not want this however since no work is produced in that case.

\textbf{Phase O}: Substituting $\xi = \braket{(k_V-k_U)^2}$ and $\chi = \braket{k_V-k_U}^2$ we find that the supremum is attained for finite $\tau^* \geq 0$ if and only if
\begin{align}
     \label{eq:Phase_Stoch_kV_Final}
     \begin{cases}
        m > 1, \\
       C_N \geq 2+N(1-m^2)\kappa 
    \end{cases}.
\end{align}
Here we have introduced $\kappa = \frac{\braket{(k_V-k_U)^2}-\braket{(k_V-k_U)}^2}{\braket{(k_V-k_U)^2}}$ for brevity. Thus the variable $\kappa$ quantifies the impact of the stochasticity in $k_V$ on the phase regions. Note that $0\leq \kappa \leq 1$ is the full range of possible values.

\textbf{Phase B}: For the other phase, when the supremum is not attained for finite $\tau^* \geq 0$ but as $\tau^* \to \infty$, we simply compute the complement to Eq. \eqref{eq:Phase_Stoch_kV}. We find that the supremum is not attained for any finite $\tau^* \geq 0$ if and only if
\begin{align}
     \label{eq:App_Phase_2_Stochastic_kV}
        m \leq 1, \;  
       C_N \leq 2+N(1-m^2)\kappa
\end{align}
\subsubsection{Phases Infinite N}
When we take $N \to \infty$ in Eq. \eqref{eq:Phase_Stoch_kV_Final}, we have many choices. If $\delta -m^2 \neq (1-m^2)\kappa $ then taking $N \to \infty$ in Eq. \eqref{eq:Phase_Stoch_kV_Final} gives us the constraint $\delta -m^2 > (1-m^2)\kappa$. While if $\delta -m^2 = (1-m^2)\kappa$ then taking $N \to \infty$ in Eq. \eqref{eq:Phase_Stoch_kV_Final} gives us $\gamma -m^2 \geq 2+(1-m^2)\kappa$. 

\textbf{Phase O} is therefore found if and only if
\begin{align}
    \label{eq:App_Phase_Stoch_kV_NInfinity}
    \begin{cases}
        m>1, \\
        \delta -m^2 > (1-m^2)\kappa \\
        \gamma \geq 2+\delta , \; \delta = m^2 +(1-m^2)\kappa 
    \end{cases}
\end{align}
To get Phase B we take $N \to \infty$ in Eq. \eqref{eq:App_Phase_2_Stochastic_kV}. If $\delta-m^2 \neq (1-m^2)\kappa$ we get the condition $\delta-m^2 < (1-m^2)\kappa$ and $m \leq 1$. If instead $\delta-m^2 = (1-m^2)\kappa$, as we take $N \to \infty$ we get the conditions $\gamma \leq 2+\delta$ and $m \leq 1$. 

\textbf{Phase B} is thus attained if and only if
\begin{align}
    \label{eq:App_Phase_2_Stoch_kV_NInfinity}
    \begin{cases}
        \delta -m^2 < (1-m^2)\kappa & m \leq 1, \\
        \gamma\leq 2 +\delta & m \leq 1, \; \delta -m^2 = (1-m^2)\kappa 
    \end{cases}.
\end{align}
\section{Plots}
In this section of the appendix, we include more plots than we did in the paper. We also discuss some technicalities of the plots such as the excluded regions.
\subsection{Excluded Region}
Observe that in all plots in this paper, the excluded region is not shown in any colour. The excluded region is the region where any of the parameters $(m, \gamma, \delta)$ have mathematically unattainable values. The region of allowed values is defined by
\begin{align}
    \label{eq:Allowed_Region}
    \mathcal{A}_N = \{(m,\gamma, \delta)\;|\;\gamma \geq m^2,\; m > 0,\; 2m^2-\gamma \leq \delta \leq \gamma,\; N(\gamma -m^2)+N(N-1)(\delta -m^2) \geq 0 \text{ and } \delta \geq 0\}\cup\{(0,0,0)\}.
\end{align}
These conditions follow from the definitions $m = \braket{x_{01}^2}$, $\gamma =\braket{x_{01}^4}$ and $\delta = \braket{x_{01}^2x_{02}^2}$. For example, we find that $\braket{|\bar{x}|^4} -\braket{|\bar{x}|^2}^2\geq 0$ implies $N(\gamma -m^2)+N(N-1)(\delta -m^2) \geq 0$ while $-(\braket{x_{01}^4}-\braket{x_{01}^2}^2)\leq \braket{x_{01}^2x_{02}^2}-\braket{x_{01}^2}^2 \leq \braket{x_{01}^4}-\braket{x_{01}^2}^2$ implies $2m^2-\gamma \leq \delta \leq \gamma$. Similarly $\braket{x_{01}^4} \geq \braket{x_{01}^2}^2$ implies $\gamma \geq m^2$ while $\braket{x_{01}^2} \geq 0$ implies $m \geq 0$.

Note also that when $m = 0$ we must also have $\gamma = 0$ and $\delta = 0$ since $m = 0$ would mean $\braket{x_{01}^2} = 0$ and hence $x_{01} = 0$. This explains the last point $(0, 0, 0)$ included in eq. \eqref{eq:Allowed_Region}.
\subsubsection{Excluded Region Infinite N}
We happen to get a different admissibility region as $N \to \infty$.

If we take $N \to \infty$ in Eq. \eqref{eq:Allowed_Region}, we see that we get the requirement that $\delta \geq m^2$, meaning that the admissible region reads
\begin{align*}
    \mathcal{A}_{\infty} = \{(m,\gamma, \delta)\; |\;\gamma \geq m^2,\; m > 0,\; 2m^2-\gamma \leq \delta \leq \gamma,\; \text{ and } \delta \geq m^2\}\cup \{(0, 0,0) \}.
\end{align*}
But since $\gamma \geq m^2$ we see that $2m^2-\gamma \leq m^2 \leq \delta$ always holds, meaning that our admissibility condition is
\begin{align}
    \mathcal{A}_{\infty} = \{(m,\gamma, \delta)\; |\;m > 0, \;  \gamma \geq \delta \geq m^2\}\cup \{(0, 0, 0) \}.
\end{align}
\subsection{Stochastic time}
\begin{figure}
    \centering
    \includegraphics[width=0.75\linewidth]{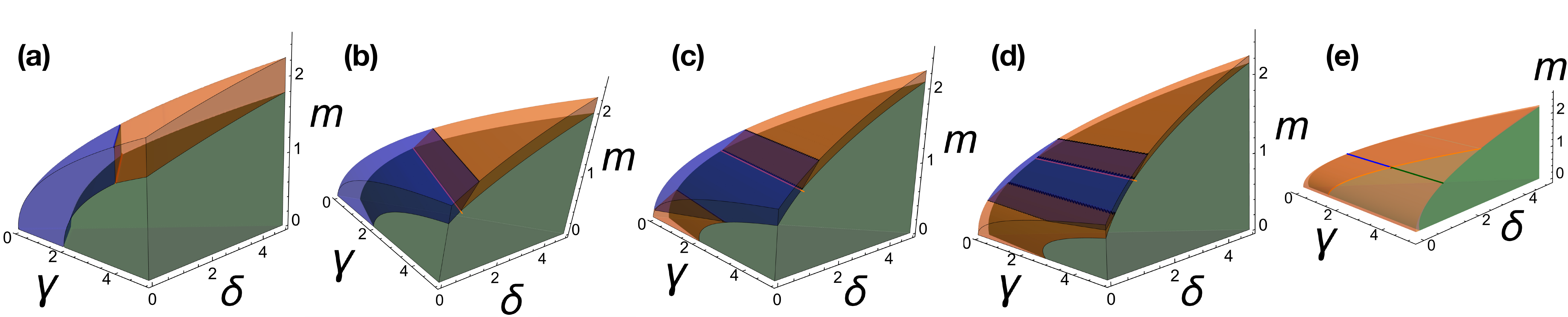}
    \caption{ The phase space structure in the stochastic $t^*$ case (a)  $N = 2$ (b) $N = 4$, (c) $N = 7$, and (d) $N = 14$. (e) $N = \infty$. In all these plots, we focus on the region $0 \leq m \leq 2.5$, $0 \leq \gamma \leq 5$ and $0 \leq \delta \leq 5$.}
    \label{fig:App_StochTime}
\end{figure}

\begin{figure}
    \centering
    \includegraphics[width=0.75\linewidth]{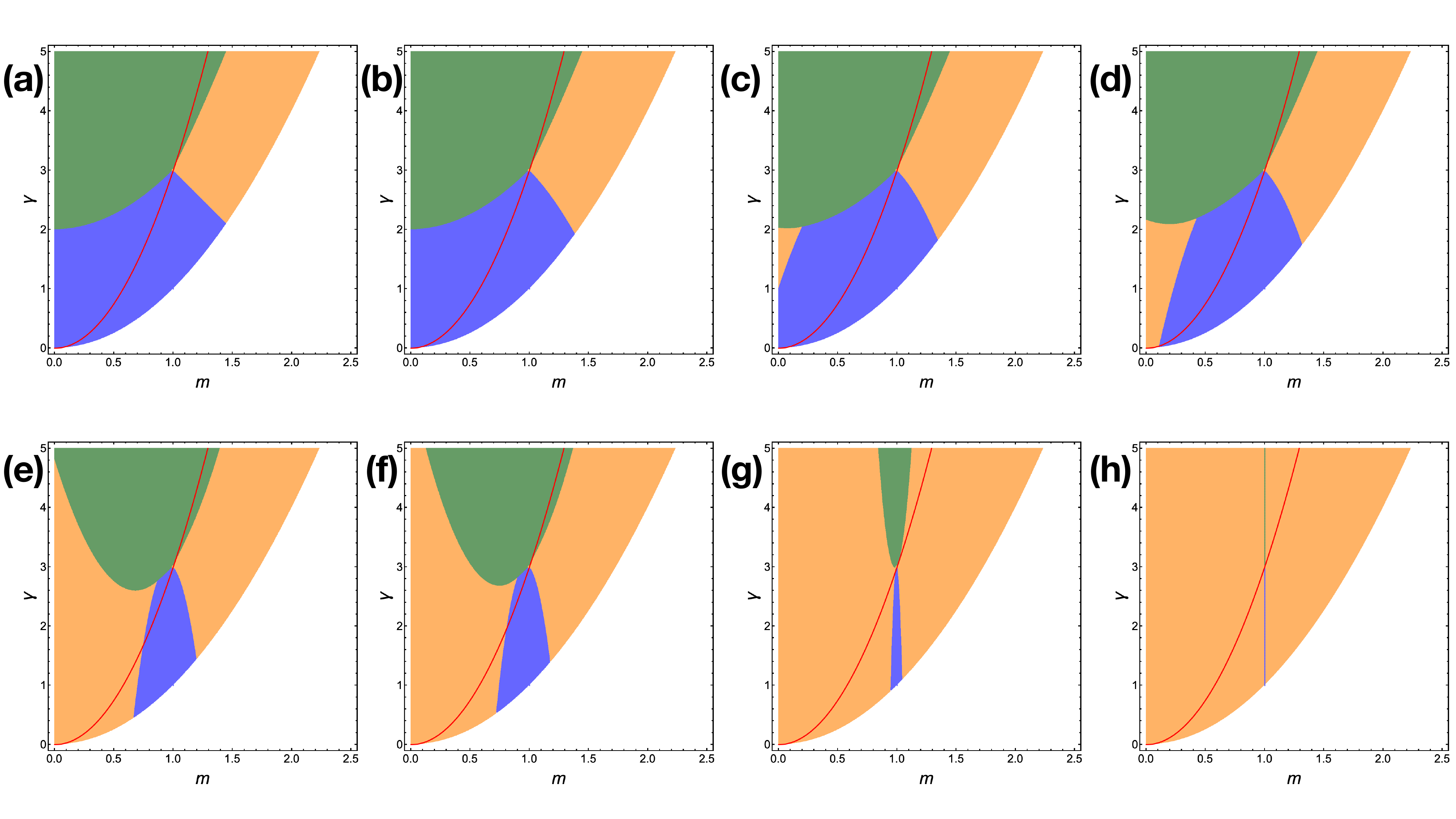}
    \caption{ A two-dimensional slice of the corresponding $3$-d phase space obtained by imposing the constraint $\delta = m^2$. All points in this $2$-d region hence correspond to factorized initial conditions. In addition the points along the red curve correspond to Gaussian initial single particle states $\gamma = 3m^2$  (a)  $N = 1$ (b)  $N = 3$, (c)  $N = 5$, and (d) $N = 7$. (e)  $N = 30$ (f)  $N = 40$ (g)  $N = 800$ (h) $N = \infty$}
    \label{fig:App_StochTimeF}
\end{figure}

\begin{figure}
    \centering
    \includegraphics[width=0.75\linewidth]{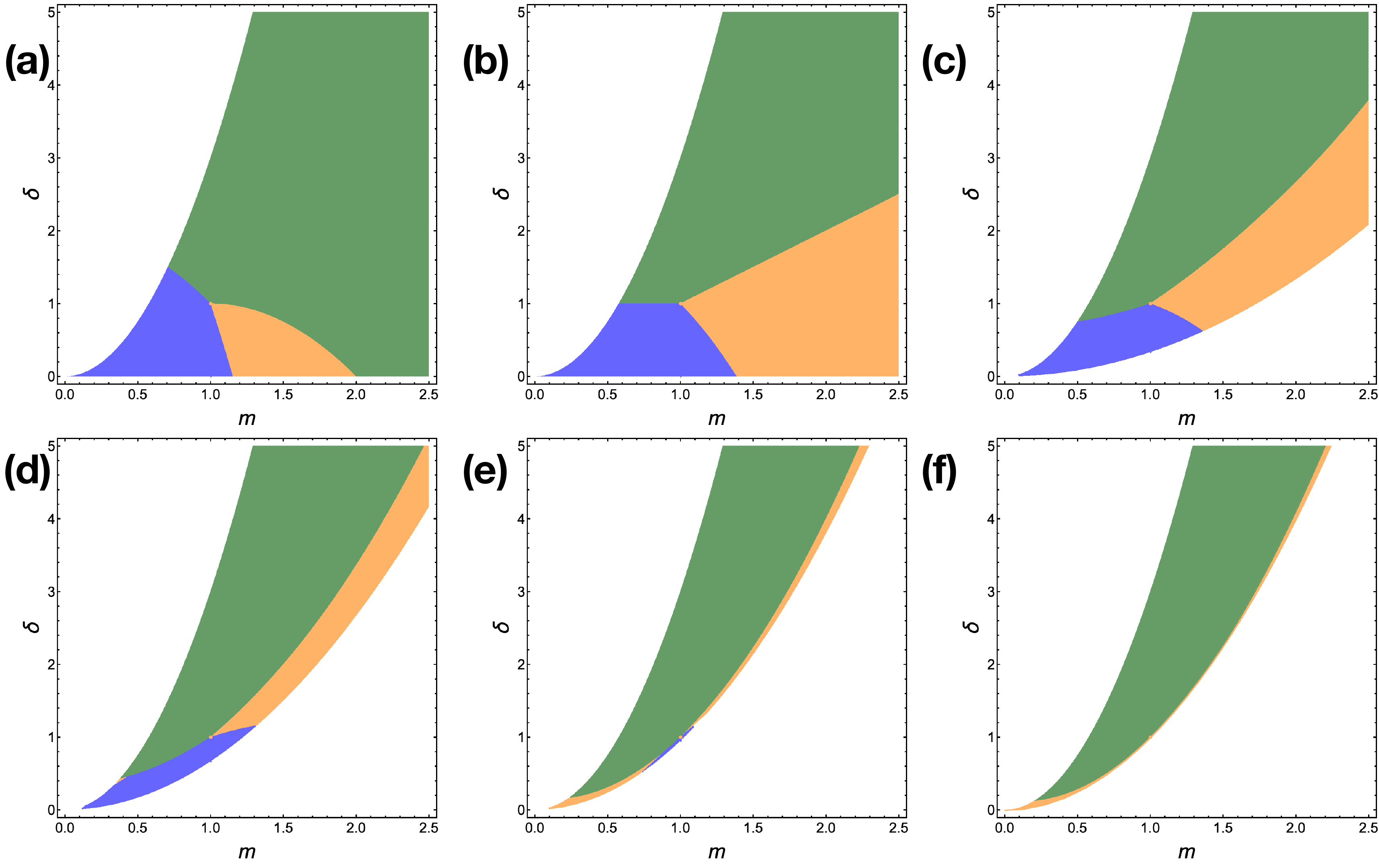}
    \caption{ All points in this set of $2$-d phase-space sections correspond to Gaussian initial single-particle states $\gamma = 3m^2$. Note though that unless $\delta = m^2$, the initial states are not factorized  (a)  $N = 2$ (b)  $N = 3$, (c)  $N = 4$, and (d)  $N = 7$. (e)  $N = 400$ and (f)  $N = \infty$. The orange dot at $(m, \delta) = (1, 1)$ corresponds to starting from an equilibrium state.} 
    \label{fig:App_StochTimeG}
\end{figure}

In this subsection, we provide the plots when the time is treated as stochastic. For finite N, we plot the conditions in Eq. \eqref{eq:App_Stoch_Time_Phase_1} in orange, the conditions in Eq. \eqref{eq:App_Stoch_Time_Phase_2A_Finite} in blue and lastly the conditions in Eq. \eqref{eq:App_Stoch_Time_Phase_2B_Finite} in green. Varying $N = 2,\; N= 4,\; N = 7$ and $N = 14$ we find (a), (b), (c) and (d) respectively in Figure \ref{fig:App_StochTime}. If $N = \infty$ we need to plot the conditions in Eq. \eqref{eq:App_Stoch_Time_Phase_1_Infinite_N} in orange to get Eq. \eqref{eq:App_Stoch_Time_Phase_2A_Infinite} in blue and Eq. \eqref{eq:App_Stoch_Time_Phase_2B_Infinite} in green to get (e) in Figure \ref{fig:App_StochTime}. If we impose $\delta = m^2$ on top of these conditions, we get Figure \ref{fig:App_StochTimeF}. When we impose $\gamma = 3m^2$ we get Figure \ref{fig:App_StochTimeG}.

\subsection{Stochastic $k_V$}
\begin{figure}
    \centering
    \includegraphics[width=0.75\linewidth]{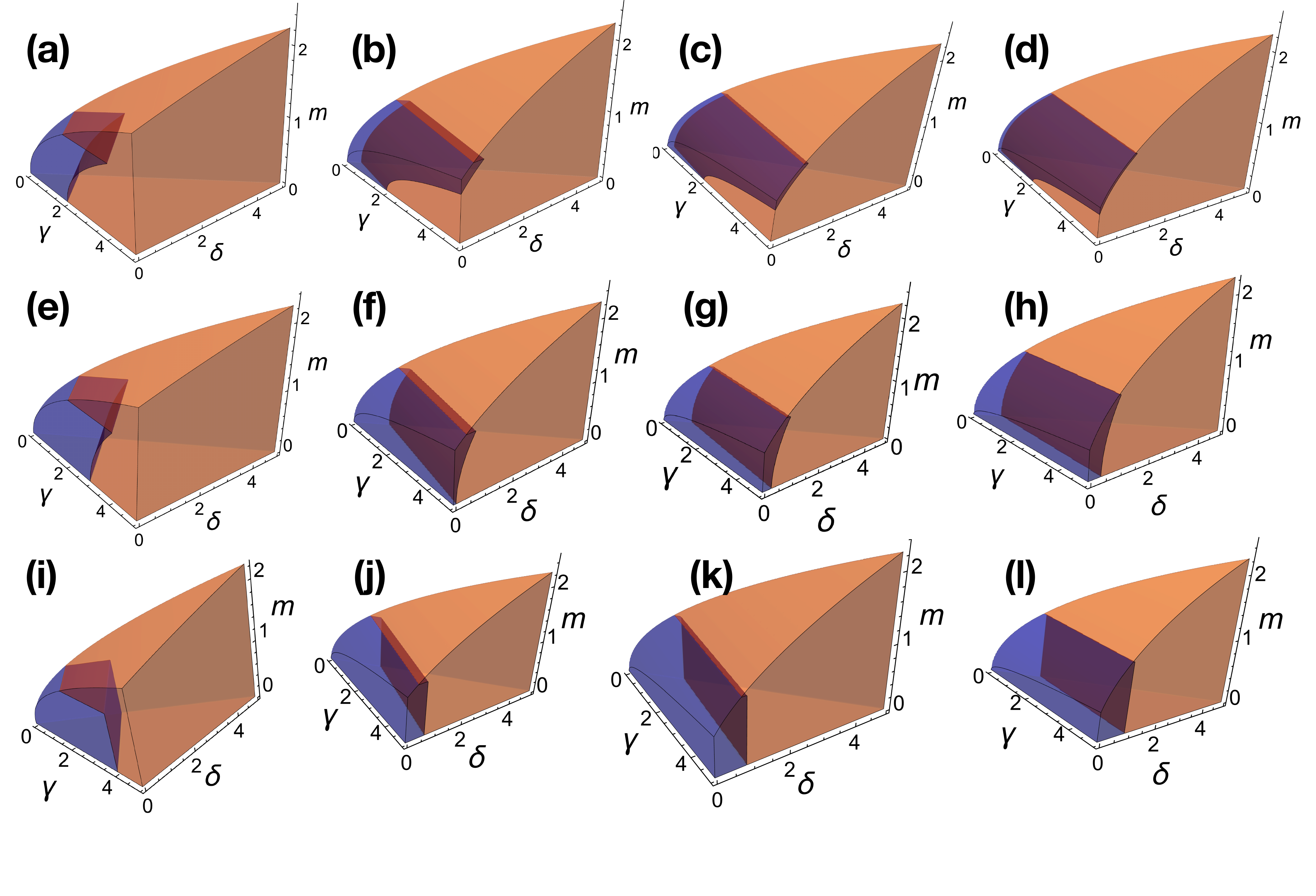}
    \caption{ Stochastic $k_V$ case for (a) $N =2$, (b) $N =4$, (c) $N = 7$ and (d) $N = 30$ all with $\kappa = 0$. (e) $N =2$, (f) $N =4$, (g) $N = 7$ and (h) $N = 30$ all with $\kappa = \frac{1}{2}$. (i) $N =2$, (j) $N =4$, (k) $N = 7$ and (l) $N = 30$ all with $\kappa = 1$. In all these plots, we focus on the region $0 \leq m \leq 2.5$, $0 \leq \gamma \leq 5$ and $0 \leq \delta \leq 5$ for the purpose of presentation.}
    \label{fig:App_StochkV_Finite_N}
\end{figure}

\begin{figure}
    \centering
    \includegraphics[width=0.75\linewidth]{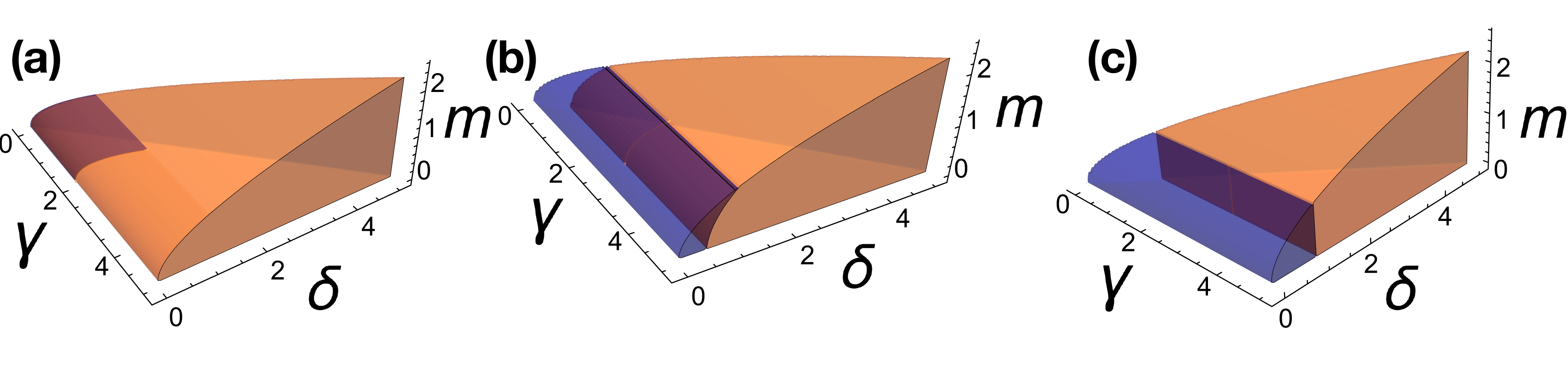}
    \caption{ Stochastic $k_V$ and $N = \infty$ for (a) $\kappa = 0$, (b) $\kappa = \frac{1}{2}$ and (c) $\kappa = 1$. Like in Figure \ref{fig:App_StochkV_Finite_N}, we focus on $0 \leq m \leq 2.5$, $0 \leq \gamma \leq 5$ and $0 \leq \delta \leq 5$ for the purpose of presentation. Note that in (a) there is a tiny sheet of Phase B on top of the orange which is roughly below $\gamma = 2$. }
    \label{fig:App_StochkV_Infinite_N}
\end{figure}

Here we consider  $k_V$ stochastic and $N$ finite. For finite $N$, we plot the conditions in Eq. \eqref{eq:Phase_Stoch_kV_Final} in orange and Eq. \eqref{eq:App_Phase_2_Stochastic_kV} in blue (Fig. \ref{fig:App_StochkV_Finite_N}). For infinite $N$ we plot the conditions in Eq. \eqref{eq:App_Phase_Stoch_kV_NInfinity} in orange and the conditions in Eq. \eqref{eq:App_Phase_2_Stoch_kV_NInfinity} in blue (Fig. \ref{fig:App_StochkV_Infinite_N}). 

\begin{figure}
    \centering
    \includegraphics[width=0.75\linewidth]{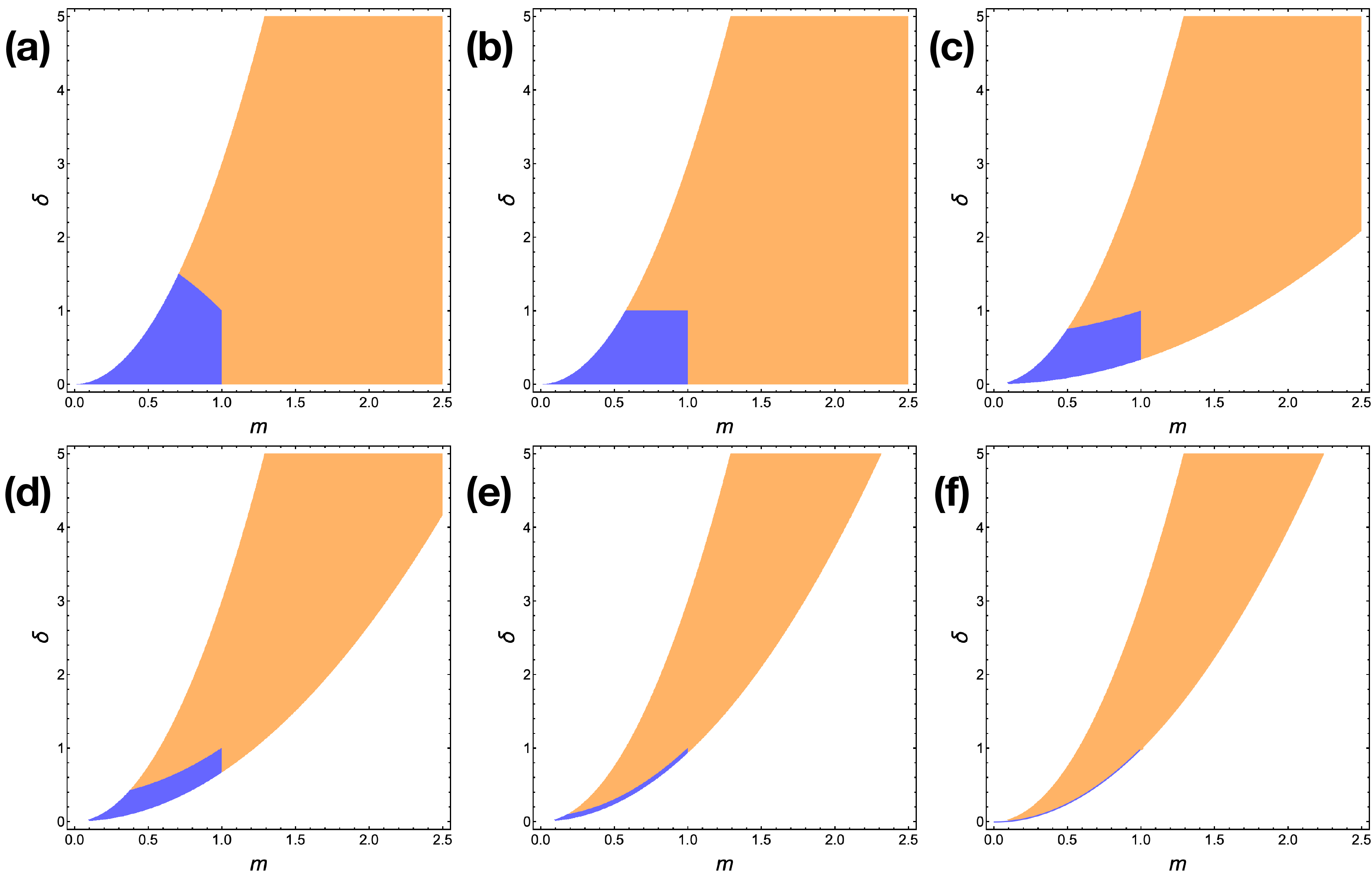}
    \caption{Here $\kappa = 0$ and $\gamma = 3m^2$. Also (a) $N = 2$, (b) $N = 3$, (c) $N = 4$, (d) $N = 7$, (e) $N = 30$, (f) $N = \infty$. Observe that in Figure \ref{fig:App_StochkVK0G}f) there is a thin curve in blue when $m < 1$}
    \label{fig:App_StochkVK0G}
\end{figure}

\begin{figure}
    \centering
    \includegraphics[width=0.75\linewidth]{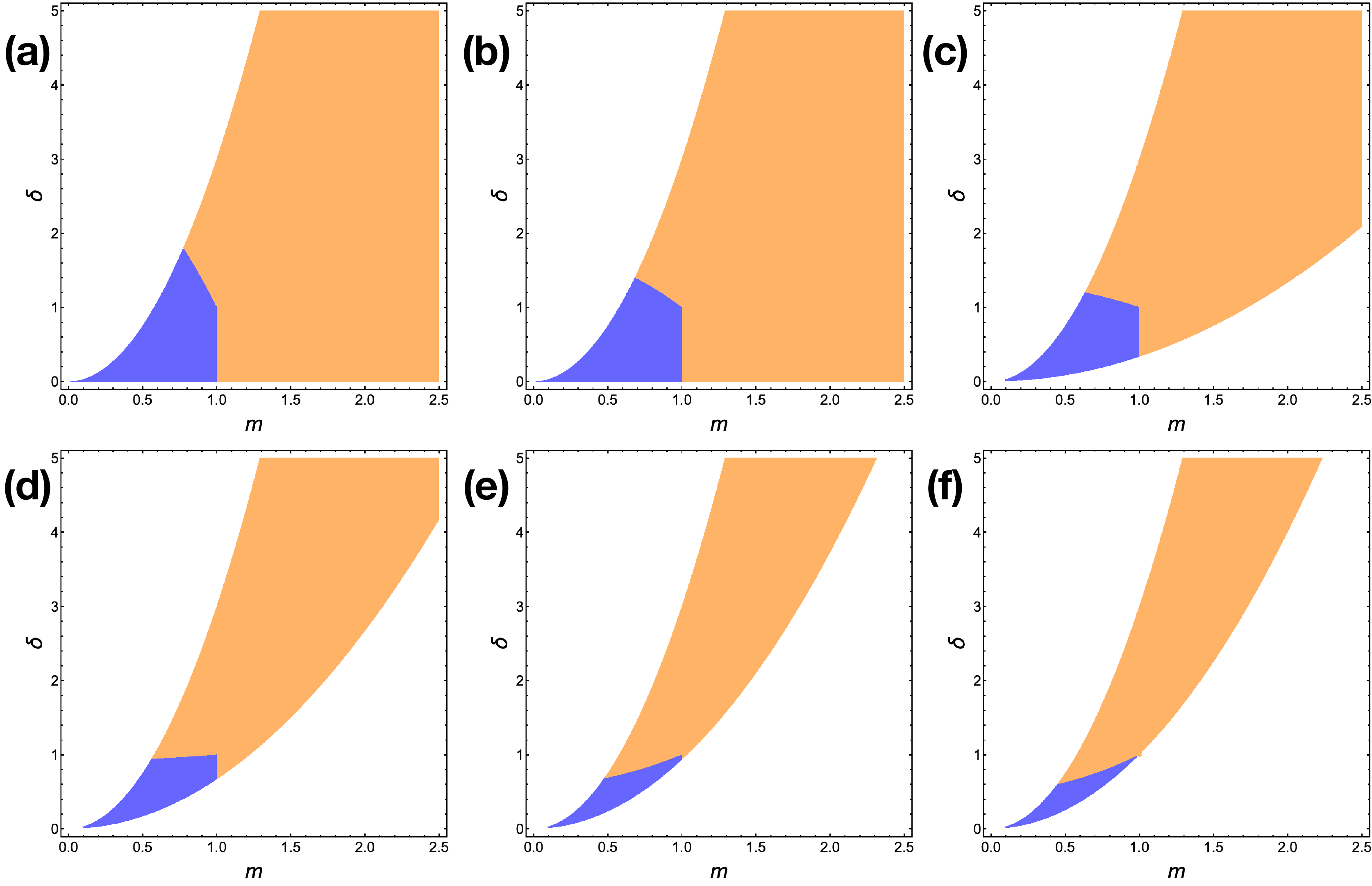}
    \caption{Here $\kappa = \frac{1}{2}$ and $\gamma = 3m^2$. Also (a) $N = 2$, (b) $N = 3$, (c) $N = 4$, (d) $N = 7$, (e) $N = 30$, (f) $N = \infty$. In contrast with Figure \ref{fig:App_StochkVK0G}f), in Figure \ref{fig:App_StochkVK0P5G} the blue is not a thin curve but has a bulk.}
    \label{fig:App_StochkVK0P5G}
\end{figure}

\begin{figure}
    \centering
    \includegraphics[width=0.75\linewidth]{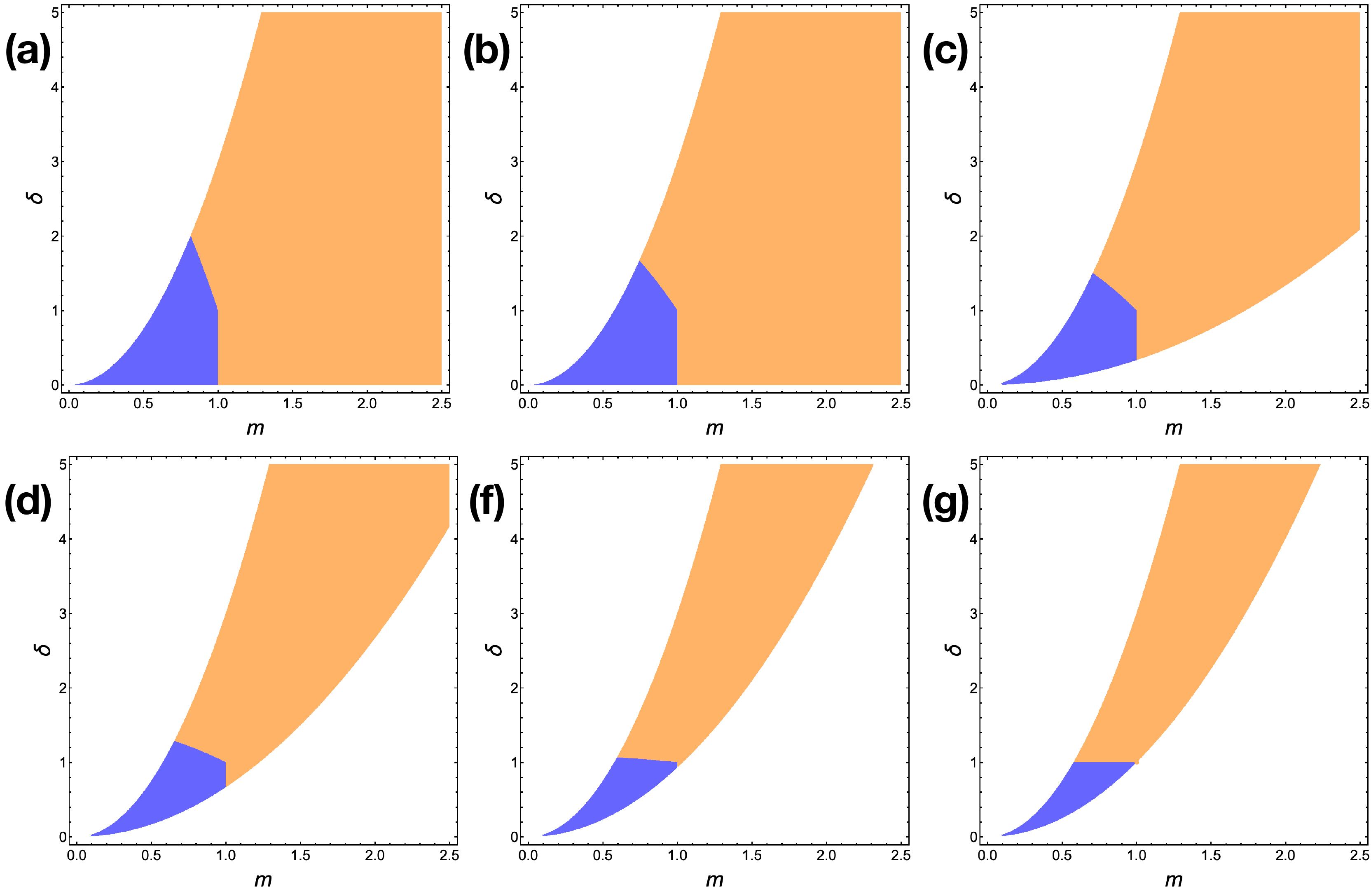}
    \caption{Here $\kappa = 1$ and $\gamma = 3m^2$. Also (a) $N = 2$, (b) $N = 3$, (c) $N = 4$, (d) $N = 7$, (e) $N = 30$, (f) $N = \infty$}
    \label{fig:App_StochkVK1G}
\end{figure}

\begin{figure}
    \centering
    \includegraphics[width=0.75\linewidth]{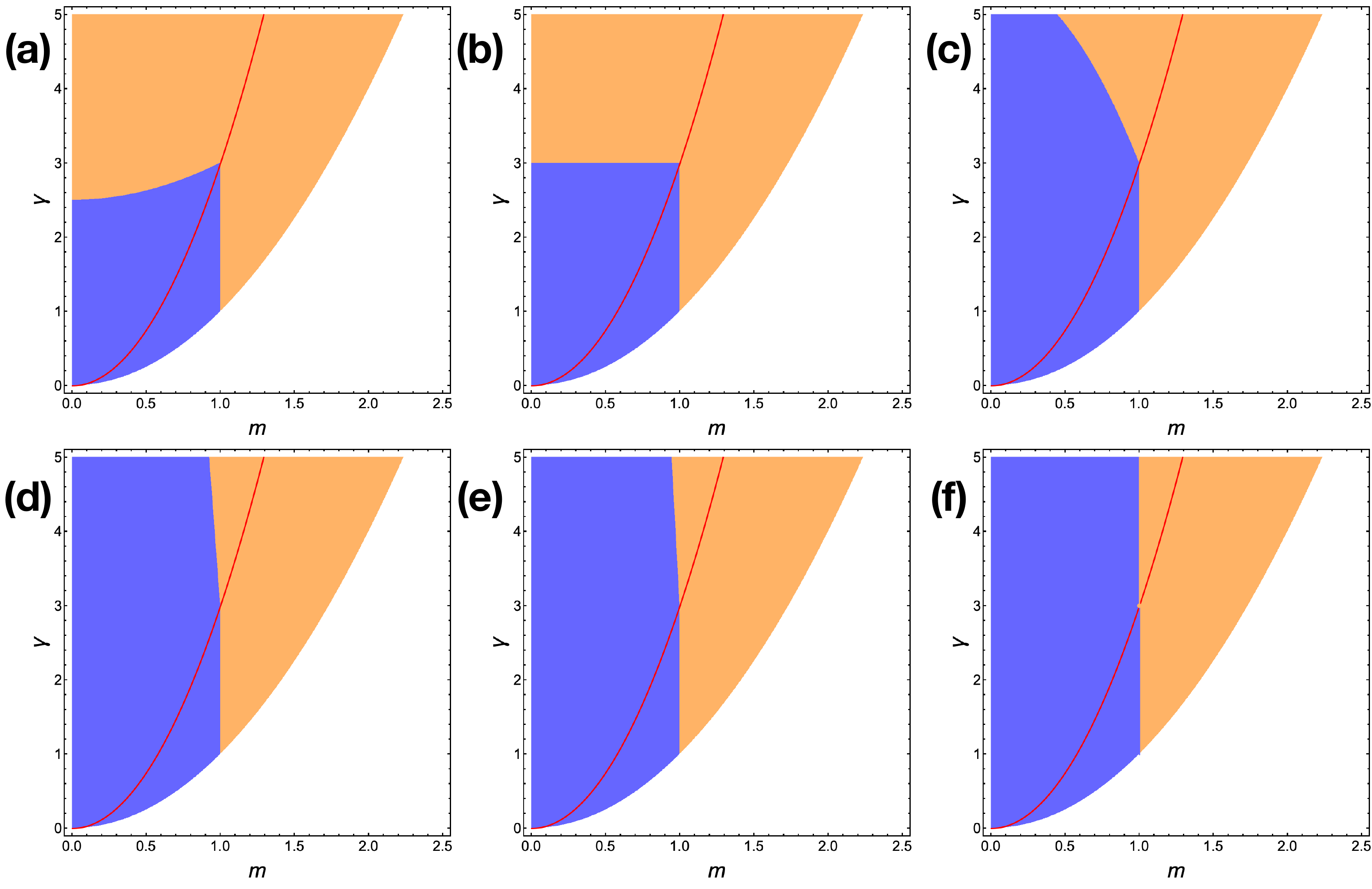}
    \caption{Here $\kappa = \frac{1}{2}$ and $\delta = m^2$. Also (a) $N = 1$, (b) $N = 2$, (c) $N = 7$, (d) $N = 30$, (e) $N = 40$, (f) $N = \infty$. Observe that in Figure \ref{fig:App_StochkVK0P5F}f) when $m = 1$ we get the Phase B when $\gamma < 3$ while we get Phase O when $\gamma \geq 3$.}
    \label{fig:App_StochkVK0P5F}
\end{figure}

\begin{figure}
    \centering
    \includegraphics[width=0.75\linewidth]{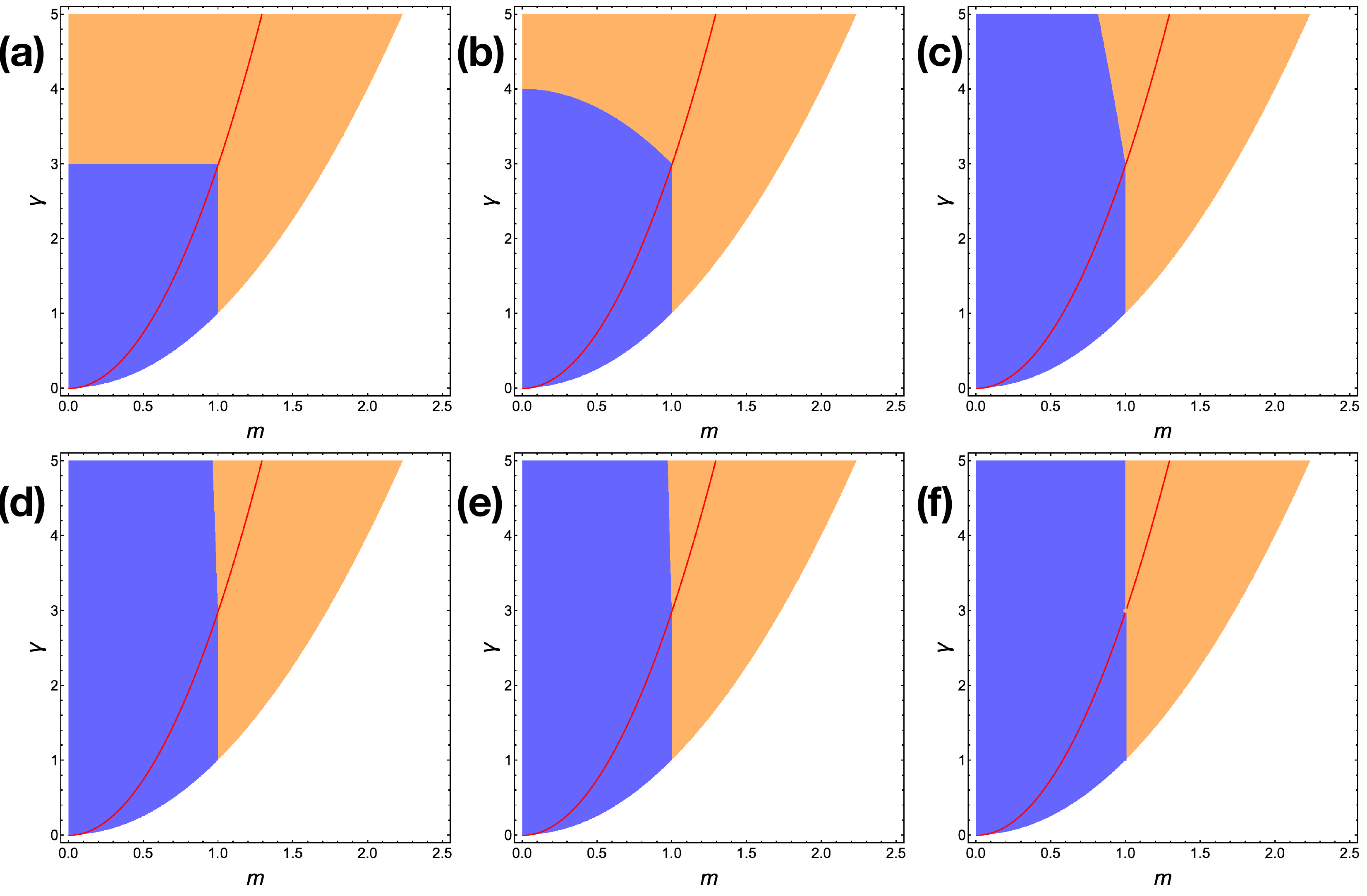}
    \caption{Here $\kappa = 1$ and $\delta = m^2$. Also (a) $N = 1$, (b) $N = 2$, (c) $N = 7$, (d) $N = 30$, (e) $N = 40$, (f) $N = \infty$. Note that Figure \ref{fig:App_StochkVK1F}f) here is the same Figure \ref{fig:App_StochkVK0P5F}f). In fact, we get the figure in (f) for $N = \infty$ whenever $0 <\kappa \leq 1$.}
    \label{fig:App_StochkVK1F}
\end{figure}

\begin{figure}
    \centering
    \includegraphics[width=0.3\linewidth]{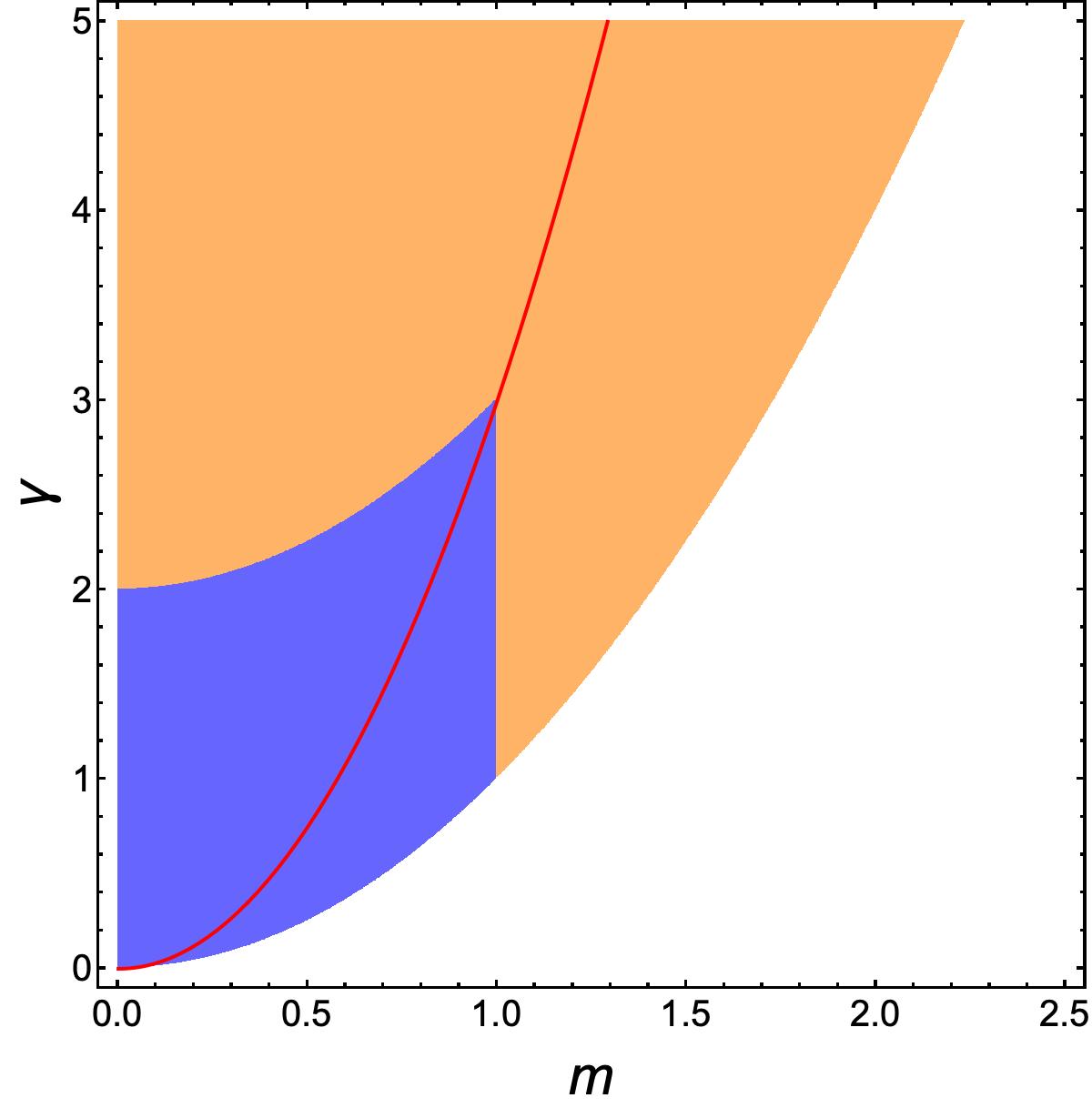}
    \caption{Whenever $\kappa = 0$ and $\delta = m^2$ we get this figure independently of $N$. That is $N = 1, 2, 3, \dots$ and $N = \infty$ all give the same plot. }
    \label{fig:StochkVK0N1F}
\end{figure}

If we restrict ourselves to $\gamma = 3m^2$ we get Figures \ref{fig:App_StochkVK0G}, \ref{fig:App_StochkVK0P5G} and \ref{fig:App_StochkVK1G}. If we instead restrict to $\delta = m^2$ we get Figures \ref{fig:App_StochkVK0P5F}, \ref{fig:App_StochkVK1F} and \ref{fig:StochkVK0N1F}.
\section{Analysis of Phase Discontinuity}
\subsection{Stochastic Time}
Here, we visualize what the phase discontinuity looks like when time $t^*$ is stochastic.

\begin{figure}
    \centering
    \includegraphics[width=0.7\linewidth]{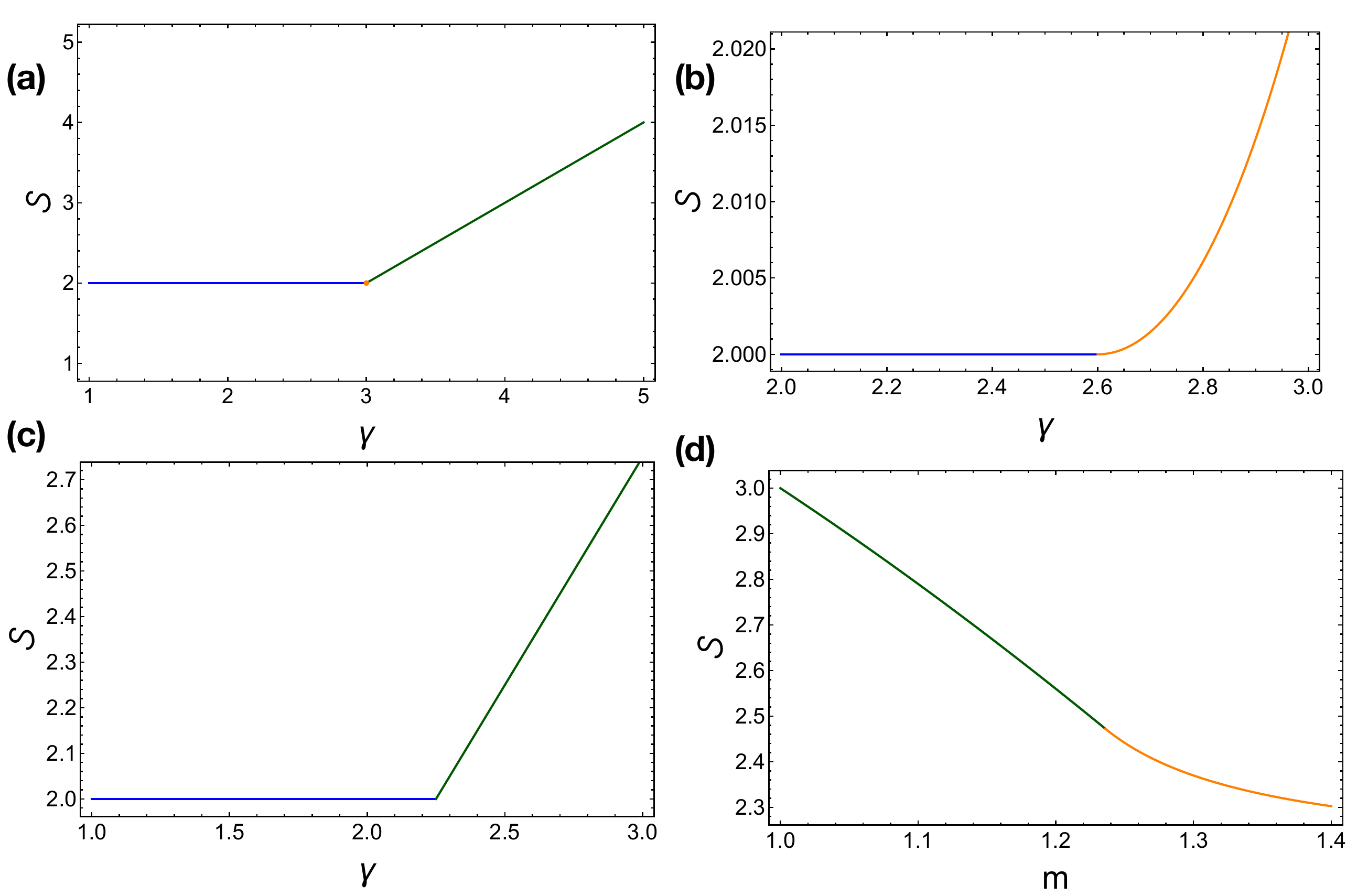}
    \caption{These plots illustrate $\mathcal{S} = \sup_{s > 0}\sigma_W^2$ plot as a function of $m$ or $\gamma$ when the other variables are fixed. The colours correspond to the phases, namely green implies that the supremum is approached as $s \to \infty$,  blue implies that the supremum is approached as $s \to 0^+$ and orange implies that the  supremum is attained for finite $s > 0$. In all these figures we have $N = 1$ and $\delta = m^2$, so the phase plots we are considering are described by Figure \ref{fig:App_StochTimeF}a). For Figure \ref{fig:StochTimeDisc}a) we have $m = 1$ and vary $1 \leq \gamma \leq 5$. In Figure \ref{fig:StochTimeDisc}b) we have $m = 1.2$ and vary $2 \leq \gamma \leq 3$. For Figure \ref{fig:StochTimeDisc}c) we have $m = 0.5$ and vary $1 \leq \gamma \leq 3$. Lastly for Figure \ref{fig:StochTimeDisc}d) we have $\gamma = 4$ and vary $1 \leq m \leq 1.4$. }
    \label{fig:StochTimeDisc}
\end{figure}

\begin{figure}
    \centering
    \includegraphics[width=0.7\linewidth]{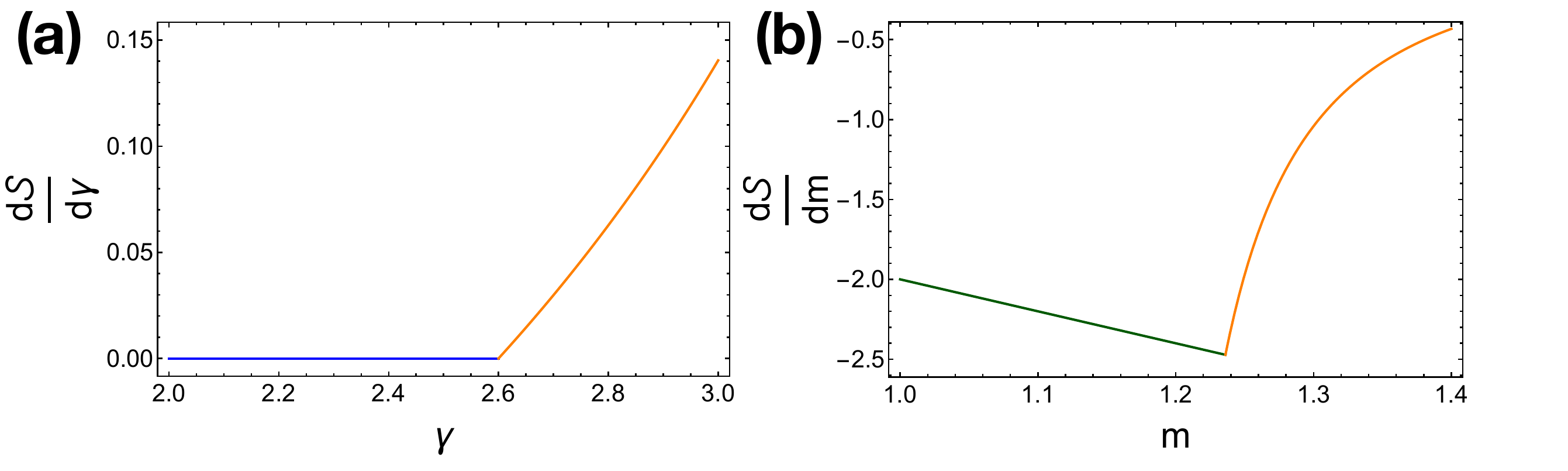}
    \caption{Plot of the derivatives of Figures \ref{fig:StochTimeDisc}b) and \ref{fig:StochTimeDisc}d) in \ref{fig:StochTimeDiscDer}a) and \ref{fig:StochTimeDiscDer}b) respectively. Otherwise the same conventions hold as in Figure \ref{fig:StochTimeDisc}. }
    \label{fig:StochTimeDiscDer}
\end{figure}

We will consider the case when $m = 1$ and $\delta = 1$ in eq. \eqref{eq:Stoch_Time_Work_Var_Nondim}. In this case, we find $\mathcal{V}(s) 
    = 2N+N\frac{s}{s+4}(\gamma-3)$. Note that when $\gamma > 3$ the supremum is attained as $s \to \infty$ while when $\gamma < 3$ it is attained as $s \to 0^+$ ($\gamma = 3$ is an equilibrium initial condition in this case with the work variance  independent of $s$). The supremum of the work variance hence has the form:
\begin{align}
    \label{eq:Supremum_Work_Var_Stoch_Time}
    \mathcal{S}^{(t)}(\gamma) = \sup_{s > 0}\sigma_W^2 = \begin{cases}
        N(\gamma - 1), & \gamma > 3 \\
        2N & 1 \leq \gamma \leq 3
    \end{cases}
\end{align}
Here, we therefore expect $\frac{\text{d}\mathcal{S}^{(t)}}{\text{d}\gamma}$ to be discontinuous. $N = 1$ and the interval $5 \geq \gamma \geq 1$ is plotted in Figure \ref{fig:StochTimeDisc}a). Figure \ref{fig:StochTimeDisc} also shows other examples of phase discontinuities that can happen.
Note that in Figures \ref{fig:StochTimeDisc}b) and \ref{fig:StochTimeDisc}d) , the phase transition is like a  second order transition. We can see this when we plot the corresponding derivatives in Figures \ref{fig:StochTimeDiscDer}a) and \ref{fig:StochTimeDiscDer}b) respectively. The derivatives are continuous but the second derivatives are not. 
\subsection{Stochastic $k_V$}
In this  section, we want to visualize how the phase discontinuity looks like when $k_V$ is stochastic. We start with Eq. \eqref{eq:Quadratic_Phase_Trans_Stoch_kV} which is the non-dimensionalized work variance when the $k_V$ is stochastic.

We will focus on the case when $\bar{A}_N < 0$. The reason is that then, Eq. \eqref{eq:Quadratic_Phase_Trans_Stoch_kV} is concave and has a unique global maximum for all $z \in \mathbb{R}$, the maximum being located at $\tilde{z} = -\frac{\bar{B}_N}{2\bar{A}_N}$. If also $1 \geq \tilde{z} >0$ then we know that the supremum is attained for finite $\tau^* \geq 0$ and the value is $\tilde{\mathcal{V}}(z = \tilde{z}) = \bar{C}_N-\frac{1}{4}\frac{\bar{B}_N^2}{\bar{A}_N}$. If $\tilde{z} > 1$ then the supremum is attained at $z = 1$ and has value $\tilde{\mathcal{V}}(z = 1) = \bar{C}_N+\bar{A}_N+\bar{B}_N$. If instead $\tilde{z} \leq 0$ then the supremum is not attained for any finite $\tau^* \geq 0$. Then the supremum value is attained as $\tau^* \to \infty$ and is given by $\tilde{\mathcal{V}}(z = 0) = \bar{C}_N$. So summarizing, we have
\begin{align}
    \label{eq:Supremum_Work_Var_Stoch_kV}
    \mathcal{S}(m) = \sup_{\tau^* \geq 0}\sigma_{W^{\text{Tot}}_N}^2= \begin{cases}
        \bar{C}_N+\bar{B}_N+\bar{A}_N,& -\frac{\bar{B}_N}{2\bar{A}_N} > 1 \\
        \bar{C}_N-\frac{1}{4}\frac{\bar{B}_N^2}{\bar{A}_N}, & 1 \geq -\frac{\bar{B}_N}{2\bar{A}_N} > 0 \\
        \bar{C}_N & -\frac{\bar{B}_N}{2\bar{A}_N} \le 0
    \end{cases}
\end{align}
Recall that here $\bar{A}_N = \frac{N}{4}\Big(\big(2+N-2m(2+N)+\gamma +(N-1)\delta\big)\xi-(m-1)^2N\chi\Big)$, $\bar{B}_N = \frac{N}{2}(m-1)\big((N+2)\xi-N\chi \big)$ and $\bar{C}_N = \frac{N}{4}\big((N+2)\xi-N\chi\big)$.

\begin{figure}
    \centering
    \includegraphics[width=0.7\linewidth]{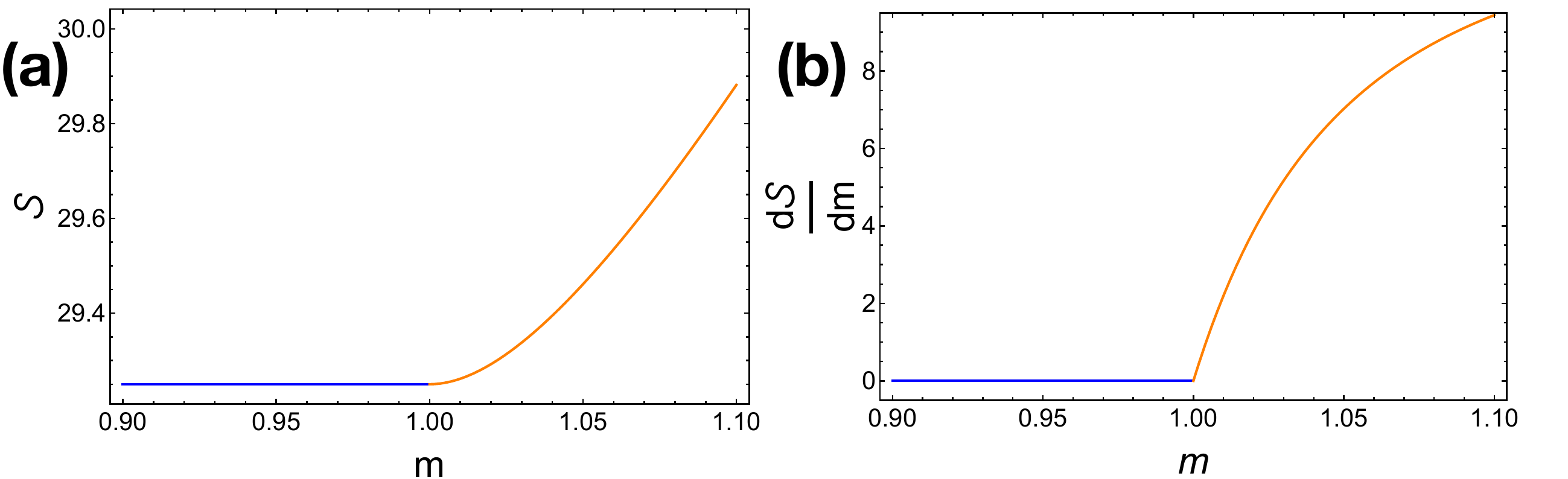}
    \caption{Plot of the phase discontinuity when $k_V$ is stochastic in Figure \ref{fig:StochkVDiscPhases}a) while its derivative is plotted in Figure \ref{fig:StochkVDiscPhases}b). Clearly, the phase transition here is second order. In the blue region, the supremum is approached as $\tau^* \to \infty$ while in the orange region it is attained for finite $\tau^* \geq 0$. }
    \label{fig:StochkVDiscPhases}
\end{figure}

We can plot eq. \eqref{eq:Supremum_Work_Var_Stoch_kV} for $\gamma = 2$, $\delta = 1$, $\xi = 9$, $\chi = 2$, $N = 3$ and when $0.9 \leq m \leq 1.1$. One can confirm that in this region $\bar{A}_N < 0$ as required. The function in eq. \eqref{eq:Supremum_Work_Var_Stoch_kV} looks like
\begin{align}
    \label{eq:Supremum_Work_Var_Stoch_kV_for_Plot}
    \mathcal{S}(m) =\begin{cases}
        \frac{117}{4} \; &0.9 \leq m\leq 1 \\
        \frac{117}{4}+\frac{1521(m-1)^2}{-100+8m(13+m)} & 1 < m \leq 1.1
    \end{cases}
\end{align}
Plotting this equation, we get Figure \ref{fig:StochkVDiscPhases}.

\bibliography{refs.bib}

@article{Biroli_2023,
  title = {Extreme Statistics and Spacing Distribution in a Brownian Gas Correlated by Resetting},
  author = {Biroli, Marco and Larralde, Hernan and Majumdar, Satya N. and Schehr, Gr{\'e}gory},
  journal = {Phys. Rev. Lett.},
  volume = {130},
  issue = {20},
  pages = {207101},
  numpages = {5},
  year = {2023},
  month = {May},
  publisher = {American Physical Society},
  doi = {10.1103/PhysRevLett.130.207101},
  url = {https://link.aps.org/doi/10.1103/PhysRevLett.130.207101}
}

@article{Entropy_UD,
  title = {Entropy production in systems with unidirectional transitions},
  author = {Busiello, D. M. and Gupta, A. and Maritan, A.},
  journal = {Phys. Rev. Res.},
  volume = {2},
  issue = {2},
  pages = {023011},
  numpages = {15},
  year = {2020},
  month = {Apr},
  publisher = {American Physical Society},
  doi = {10.1103/PhysRevResearch.2.023011},
  url = {https://link.aps.org/doi/10.1103/PhysRevResearch.2.023011}
}

@article{Deepak2022_work,
doi = {10.1088/1367-2630/aca25e},
url = {https://doi.org/10.1088/1367-2630/aca25e},
year = {2022},
month = {nov},
publisher = {IOP Publishing},
volume = {24},
number = {11},
pages = {113034},
author = {Gupta, Deepak and Plata, Carlos A},
title = {Work fluctuations for diffusion dynamics submitted to stochastic return},
journal = {New Journal of Physics}
}

@article{Biroli_2024,
  title = {Exact extreme, order, and sum statistics in a class of strongly correlated systems},
  author = {Biroli, Marco and Larralde, Hern{\'a}n and Majumdar, Satya N. and Schehr, Gr{\'e}gory},
  journal = {Phys. Rev. E},
  volume = {109},
  issue = {1},
  pages = {014101},
  numpages = {18},
  year = {2024},
  month = {Jan},
  publisher = {American Physical Society},
  doi = {10.1103/PhysRevE.109.014101},
  url = {https://link.aps.org/doi/10.1103/PhysRevE.109.014101}
}

@article{Biroli_2024L,
  title = {Dynamically emergent correlations between particles in a switching harmonic trap},
  author = {Biroli, Marco and Kulkarni, Manas and Majumdar, Satya N. and Schehr, Gr{\'e}gory},
  journal = {Phys. Rev. E},
  volume = {109},
  issue = {3},
  pages = {L032106},
  numpages = {6},
  year = {2024},
  month = {Mar},
  publisher = {American Physical Society},
  doi = {10.1103/PhysRevE.109.L032106},
  url = {https://link.aps.org/doi/10.1103/PhysRevE.109.L032106}
}

@article{Sabhapandit_2024,
doi = {10.1088/1751-8121/ad6651},
url = {https://doi.org/10.1088/1751-8121/ad6651},
year = {2024},
month = {aug},
publisher = {IOP Publishing},
volume = {57},
number = {33},
pages = {335003},
author = {Sabhapandit, Sanjib and Majumdar, Satya N},
title = {Noninteracting particles in a harmonic trap with a stochastically driven center},
journal = {Journal of Physics A: Mathematical and Theoretical}
}

@article{Majumdar_2026,
doi = {10.1209/0295-5075/ae6a1e},
url = {https://doi.org/10.1209/0295-5075/ae6a1e},
year = {2026},
month = {jun},
publisher = {EDP Sciences, IOP Publishing and Società Italiana di Fisica},
volume = {155},
number = {1},
pages = {11001},
author = {Majumdar, Satya N. and Schehr, Grégory},
title = {Dynamically emergent correlations},
journal = {Europhysics Letters}
}

@article{gupta2025thermodynamic,
  title={Thermodynamic cost of recurrent erasure},
  author={Gupta, Deepak and Olsen, Kristian St{\o}levik and Krishnamurthy, Supriya},
  journal={Communications Physics},
  volume={8},
  number={1},
  pages={355},
  year={2025},
  publisher={Nature Publishing Group UK, London}
}

@article{mercado2020intermittent,
  title={Intermittent resetting potentials},
  author={Mercado-V{\'a}squez, Gabriel and Boyer, Denis and Majumdar, Satya N and Schehr, Gr{\'e}gory},
  journal={Journal of Statistical Mechanics: Theory and Experiment},
  volume={2020},
  number={11},
  pages={113203},
  year={2020},
  publisher={IOP Publishing}
}

@article{santra2021brownian,
  title={Brownian motion under intermittent harmonic potentials},
  author={Santra, Ion and Das, Santanu and Nath, Sujit Kumar},
  journal={Journal of Physics A: Mathematical and Theoretical},
  volume={54},
  number={33},
  pages={334001},
  year={2021},
  publisher={IOP Publishing}
}

@article{Yael_2025,
  title = {Many-body colloidal dynamics under stochastic resetting: Competing effects of particle interactions on the steady-state distribution},
  author = {Vatash, Ron and Roichman, Yael},
  journal = {Phys. Rev. Res.},
  volume = {7},
  issue = {3},
  pages = {L032020},
  numpages = {7},
  year = {2025},
  month = {Jul},
  publisher = {American Physical Society},
  doi = {10.1103/wz1y-hgnk},
  url = {https://link.aps.org/doi/10.1103/wz1y-hgnk}
}

@article{deMauro_2026,
  title = {Dynamically emergent correlations in Brownian particles subject to simultaneous non-Poissonian resetting protocols},
  author = {de Mauro, Gabriele and Biroli, Marco and Majumdar, Satya N. and Schehr, Gr{\'e}gory},
  journal = {Phys. Rev. E},
  volume = {113},
  issue = {1},
  pages = {014120},
  numpages = {23},
  year = {2026},
  month = {Jan},
  publisher = {American Physical Society},
  doi = {10.1103/vmhz-6h5s},
  url = {https://link.aps.org/doi/10.1103/vmhz-6h5s}
}

@article{Pal_2022,
doi = {10.1088/1751-8121/ac3cdf},
url = {https://doi.org/10.1088/1751-8121/ac3cdf},
year = {2022},
month = {jan},
publisher = {IOP Publishing},
volume = {55},
number = {2},
pages = {021001},
author = {Pal, Arnab and Kostinski, Sarah and Reuveni, Shlomi},
title = {The inspection paradox in stochastic resetting},
journal = {Journal of Physics A: Mathematical and Theoretical}
}

@article{Shamik_2022,
AUTHOR={Gupta, Shamik  and Jayannavar, Arun M. },
TITLE={Stochastic Resetting: A (Very) Brief Review},
JOURNAL={Frontiers in Physics},
VOLUME={Volume 10 - 2022},
YEAR={2022},
URL={https://www.frontiersin.org/journals/physics/articles/10.3389/fphy.2022.789097},
DOI={10.3389/fphy.2022.789097},
ISSN={2296-424X}}

@article{gupta2020stochastic,
	author = {Gupta, Deepak and Plata, Carlos A and Kundu, Anupam and Pal, Arnab},
	journal = {Journal of Physics A: Mathematical and Theoretical},
	number = {2},
	pages = {025003},
	publisher = {IOP Publishing},
	title = {Stochastic resetting with stochastic returns using external trap},
	volume = {54},
	year = {2020}}

@article{mercado2022reducing,
  title={Reducing mean first passage times with intermittent confining potentials: a realization of resetting processes},
  author={Mercado-V{\'a}squez, Gabriel and Boyer, Denis and Majumdar, Satya N},
  journal={Journal of Statistical Mechanics: Theory and Experiment},
  volume={2022},
  number={9},
  pages={093202},
  year={2022},
  publisher={IOP Publishing}
}

@article{gupta2021resetting,
  title={Resetting with stochastic return through linear confining potential},
  author={Gupta, Deepak and Pal, Arnab and Kundu, Anupam},
  journal={Journal of Statistical Mechanics: Theory and Experiment},
  volume={2021},
  number={4},
  pages={043202},
  year={2021},
  publisher={IOP Publishing}
}

@article{alston2022non,
  title={Non-equilibrium thermodynamics of diffusion in fluctuating potentials},
  author={Alston, Henry and Cocconi, Luca and Bertrand, Thibault},
  journal={Journal of Physics A: Mathematical and Theoretical},
  year={2022},
  publisher={IOP Publishing}
}

@article{Biroli_2025,
  title = {Experimental Evidence for Strong Emergent Correlations between Particles in a Switching Trap},
  author = {Biroli, Marco and Ciliberto, Sergio and Kulkarni, Manas and Majumdar, Satya N. and Petrosyan, Artyom and Schehr, Gr{\'e}gory},
  journal = {Phys. Rev. Lett.},
  volume = {137},
  issue = {3},
  pages = {037102},
  numpages = {8},
  year = {2026},
  month = {Jul},
  publisher = {American Physical Society},
  doi = {10.1103/jkn8-h939},
  url = {https://link.aps.org/doi/10.1103/jkn8-h939}
}

@misc{olsen2026,
      title={Information-fluctuation inequalities for collective response}, 
      author={Kristian Stølevik Olsen},
      year={2026},
      eprint={2603.01852},
      archivePrefix={arXiv},
      primaryClass={cond-mat.stat-mech},
      url={https://arxiv.org/abs/2603.01852}, 
}

@article{tal_friedman_2020,
author = {Tal Friedman, Ofir and Pal, Arnab and Sekhon, Amandeep and Reuveni, Shlomi and Roichman, Yael},
year = {2020},
month = {08},
pages = {},
title = {Experimental Realization of Diffusion With Stochastic Resetting},
volume = {11},
journal = {The Journal of Physical Chemistry Letters},
doi = {10.1021/acs.jpclett.0c02122}
}

@article{fuchs2016stochastic,
  title={Stochastic thermodynamics of resetting},
  author={Fuchs, Jaco and Goldt, Sebastian and Seifert, Udo},
  journal={Europhysics Letters},
  volume={113},
  number={6},
  pages={60009},
  year={2016},
  publisher={IOP Publishing}
}

@article{evans2020stochastic,
  title={Stochastic resetting and applications},
  author={Evans, Martin R and Majumdar, Satya N and Schehr, Gr{\'e}gory},
  journal={Journal of Physics A: Mathematical and Theoretical},
  volume={53},
  number={19},
  pages={193001},
  year={2020},
  publisher={IOP Publishing}
}

@article{olsen2024thermodynamic,
  title={Thermodynamic cost of finite-time stochastic resetting},
  author={Olsen, Kristian St{\o}levik and Gupta, Deepak and Mori, Francesco and Krishnamurthy, Supriya},
  journal={Physical Review Research},
  volume={6},
  number={3},
  pages={033343},
  year={2024},
  publisher={APS}
}

@article{mori2023entropy,
  title={Entropy production of resetting processes},
  author={Mori, Francesco and Olsen, Kristian St{\o}levik and Krishnamurthy, Supriya},
  journal={Physical Review Research},
  volume={5},
  number={2},
  pages={023103},
  year={2023},
  publisher={APS}
}

@article{olsen2024thermodynamic2,
  title={Thermodynamic work of partial resetting},
  author={Olsen, Kristian St{\o}levik and Gupta, Deepak},
  journal={Journal of Physics A: Mathematical and Theoretical},
  volume={57},
  number={24},
  pages={245001},
  year={2024},
  publisher={IOP Publishing}
}

@article{goerlich2024resetting,
  title={Resetting as a swift equilibration protocol in an anharmonic potential},
  author={Goerlich, R{\'e}mi and Keidar, Tommer D and Roichman, Yael},
  journal={Physical Review Research},
  volume={6},
  number={3},
  pages={033162},
  year={2024},
  publisher={APS}
}

@misc{goerlich2024tamingmaxwellsdemonexperimental,
      title={Taming a Maxwell's demon for experimental stochastic resetting}, 
      author={Rémi Goerlich and Minghao Li and Luís Barbosa Pires and Paul-Antoine Hervieux and Giovanni Manfredi and Cyriaque Genet},
      year={2024},
      eprint={2306.09503},
      archivePrefix={arXiv},
      primaryClass={cond-mat.stat-mech},
      url={https://arxiv.org/abs/2306.09503}, 
}

@article{Besga_2020,
  title = {Optimal mean first-passage time for a Brownian searcher subjected to resetting: Experimental and theoretical results},
  author = {Besga, Benjamin and Bovon, Alfred and Petrosyan, Artyom and Majumdar, Satya N. and Ciliberto, Sergio},
  journal = {Phys. Rev. Res.},
  volume = {2},
  issue = {3},
  pages = {032029},
  numpages = {5},
  year = {2020},
  month = {Jul},
  publisher = {American Physical Society},
  doi = {10.1103/PhysRevResearch.2.032029},
  url = {https://link.aps.org/doi/10.1103/PhysRevResearch.2.032029}
}

@article{Faisant_2021,
doi = {10.1088/1742-5468/ac2cc7},
url = {https://dx.doi.org/10.1088/1742-5468/ac2cc7},
year = {2021},
month = {nov},
publisher = {IOP Publishing and SISSA},
volume = {2021},
number = {11},
pages = {113203},
author = {Faisant, F and Besga, B and Petrosyan, A and Ciliberto, S and Majumdar, Satya N},
title = {Optimal mean first-passage time of a Brownian searcher with resetting in one and two dimensions: experiments, theory and numerical tests},
journal = {Journal of Statistical Mechanics: Theory and Experiment}
}

\end{document}